\documentclass[fleqn,usenatbib]{mnras}

\usepackage{newtxtext,newtxmath}
\usepackage{amsmath}
\usepackage{siunitx}
\usepackage{array}
\usepackage{threeparttablex}
\usepackage{booktabs}
\usepackage{graphicx}  
\usepackage{caption}   
\usepackage{subcaption} 
\usepackage{longtable}
\usepackage{supertabular}
\usepackage{hyperref}
\usepackage[T1]{fontenc}
\usepackage{tablefootnote} 
\usepackage{threeparttable}

\DeclareRobustCommand{\VAN}[3]{#2}
\let\VANthebibliography\thebibliography
\def\thebibliography{\DeclareRobustCommand{\VAN}[3]{##3}\VANthebibliography}

\usepackage{graphicx}	
\usepackage{amsmath}	
\usepackage{ulem}

\title[MeerKAT stellar sources]{The stellar population in the SARAO MeerKAT Galactic Plane Survey}

\author[O. D. Egbo et al.]{
Okwudili D. Egbo,$^{1,2}$\thanks{E-mail: egbo@saao.ac.za (ODE)}
D. A. H. Buckley,$^{1,2,3}$
P. J. Groot,$^{1,2,4}$
F. Cavallaro,$^{5,6}$
P. A. Woudt,$^{1}$
\newauthor 
M. A. Thompson,$^{7}$
M. Mutale,$^{7}$
and M. Bietenholz$^{8,9}$ 
\\
$^{1}$Department of Astronomy, University of Cape Town, Private Bag X3, Rondebosch 7701, South Africa\\
$^{2}$South African Astronomical Observatory, PO Box 9, Observatory 7935, South Africa\\
$^{3}$Department of Physics, University of the Free State, PO Box 339, Bloemfontein 9300, South Africa\\
$^{4}$Department of Astrophysics/IMAPP, Radboud University, PO 9010, NL-6500 GL Nijmegen, the Netherlands\\
$^{5}$Universita degli studi di Catania, I-95123 Catania, Italy \\
$^{6}$INAF–Osservatorio Astrofisico di Catania, Via S. Sofia 78, I-95123 Catania, Italy \\ 
$^{7}$School of Physics and Astronomy, University of Leeds, Leeds LS2 9JT, UK \\
$^{8}$SARAO/Hartebeesthoek Radio Astronomy Observatory, PO Box 443,
Krugersdorp 1740, South Africa \\
$^{9}$Department of Physics and Astronomy, York University, Toronto, M3J 1P3
Ontario, Canada
}

\date{Accepted XXX. Received YYY; in original form ZZZ}

\pubyear{2024}

\begin{document}
\label{firstpage}
\pagerange{\pageref{firstpage}--\pageref{lastpage}}
\maketitle


\begin{abstract}

We report on optically selected stellar candidates of SARAO MeerKAT 1.3 GHz radio continuum survey sources of the Galactic plane. 
Stellar counterparts to radio sources are selected by cross-matching the MeerKAT source positions with \textit{Gaia} DR3, using two approaches. The first approach evaluated the probability of chance alignments between the radio survey and \textit{Gaia} sources and used AllWISE infrared colour-colour information to select potential stellar candidates. The second approach utilized a Monte Carlo method to evaluate the cross-matching reliability probability, based on populations of known radio-emitting stars. From the combined approaches, we found 629 potential stellar counterparts, of which 169 have existing SIMBAD classifications, making it the largest Galactic plane radio-optical crossmatch sample to date.  
A colour-magnitude analysis of the sample revealed a diverse population of stellar objects, ranging from massive OB stars, main-sequence stars, giants, young stellar objects, emission line stars, red dwarfs and white dwarfs. Some of the proposed optical counterparts include chromospherically/coronally active stars, for example RS CVn binaries, BY Dra systems, YSOs and flare stars, which typically exhibit radio emission. Based on Gaia's low-resolution spectroscopy, some of the stars show strong H$\alpha$ emission, indicating they are magnetically active, consistent with them being radio emitters. While MeerKAT's sensitivity and survey speed make it ideal for detecting faint radio sources, its angular resolution limits accurate counterpart identification for crowded fields such as the Galactic Plane. Higher frequency, and, thereby, better spatial resolution, radio observations plus circular polarization would be required to strengthen the associations.

\end{abstract}



\section{Introduction} 

Both thermal and non-thermal emission mechanisms are responsible for the radio emission detected in stars \citep{Abbott:1981, Dulk:1985, Abbott:1986, Bieging:1989, gudel2002radiostars, Matthews_2013radio, Matthews_2018radio}.
Thermal emission such as Bremsstrahlung (free-free) emission is associated with early-type massive stars and are commonly used to estimate the mass loss rates in stellar winds \citep{Stevens:1995, Montes:2009}.
In the case of non-thermal radio emission, magnetic activity in stellar atmospheres drives the relativistic (synchrotron) or mildly-relativistic (gyrosynchrotron) emission detected in active stars for both binaries and single stars \citep{Dulk:1985, Feigelson:1985, Melrose:1987, Storey:1995}. The radio emission of single main-sequence stars is largely dependent on stellar rotation, magnetic activity, and age \citep{Skumanich:1972}. For close binary systems, tidal interactions enhance stellar rotation, prevent the slowdown observed in single stars and allow for high magnetic activity levels \citep{Skumanich:1972, gudel2002radiostars, McLean:2012}. Hence, binary systems often exhibit stronger and more frequent magnetic phenomena, such as flares and coronal mass ejections, which enhance their radio emission. Recent reviews of radio stars and their emission mechanisms have been described in \citet{Seaquist:1993, gudel2002radiostars, Montes:2009, Matthews_2013radio, Umana:2015, Anglada:2018, Matthews_2018radio, Yu_2021, Callingham:2021, Driessen:2024}.

\cite{Wendker:1995} compiled a catalogue of known radio stars from the literature. \cite{gudel2002radiostars} further expanded on this work by producing a colour-magnitude diagram of 440 radio stars, describing the physics of the radio emission at different stellar evolutionary stages. These stars have been studied across a wide range of stellar systems. The massive stars, particularly in systems like OB binaries and Wolf-Rayet stars, are known for their strong stellar winds, which in turn generate shocks that accelerate particles, producing radio emission \citep{DeBecker:2013}. Main-sequence stars, many of them binaries, such as BY Dra, FK Com, and RS CVn-type systems, have also been topics of interest where their emission is non-thermal, highly variable and largely circularly polarized \citep{Owen:1978, Kuijpers:1985, Mutel:1985, Guedel:1993, Benz:2010, Toet:2021, Zhang:2022}. Low-mass stars that exhibit chromospheric and coronal activity, resulting in flares, such as dMe stars, have been explored in the context of radio emission \citep{Yu_2021, Pritchard:2021, Driessen:2022}. Young stellar objects (YSOs), including T Tauri stars, with accretion disks surrounding their young stellar cores, have emerged as interesting objects of study within the field of radio star research \citep{gudel2002radiostars, Anglada:2018}.

Radio stellar population studies have greatly benefited from recent advances in radio surveys, which have led to new catalogues of radio stars. For example, the Sydney Radio Star Catalogue (SRSC) presents a significant achievement by compiling 839 stars with 3,405 detections across MHz–GHz frequencies using the Australian SKA Pathfinder (ASKAP), mostly through circular polarization studies, with sources comprising populations of ultracool dwarfs, giants, and Wolf-Rayet stars \citep{Driessen:2024}. \cite{Pritchard:2024} identified an additional 76 stars, mostly dominated by M dwarfs, with detections arising from radio bursts.

Although normal stars are expected to be radio emitters, the limited sensitivity of radio telescopes has so far only allowed for the detection of either the most exceptionally radio-bright stars, typically closer ones.  MeerKAT's large field of view and high sensitivity have allowed us to substantially increase the number of stars detected in the radio, and its relatively high angular resolution allows deeper surveys in crowded regions like the Galactic Plane.  Taking advantage of MeerKAT's capabilities, the SARAO MeerKAT Galactic Plane Survey, hereafter SMGPS \citep{Goedhart:2024} mapped the population of radio sources in the Galactic Plane.

In this paper, we present a new radio catalogue of 629 radio candidates derived from crossmatching the SMGPS compact source catalogue  \citep{Goedhart:2024, mutale:2024} with the results from \textit{Gaia} Data Release 3 (DR3) \citep{GaiaCollaboration:2023, GaiaCollaboration:2021}. In Section 2, we describe the properties of SMGPS and \textit{Gaia} catalogues. The crossmatch methods used in identifying reliable radio-optical stellar candidates are described in Section 3.  In Section 4, we present the results of the stellar sample and a discussion on the different stellar populations in our sample.

\section{Data Description}
\subsection{SARAO MeerKAT Galactic Plane Survey}

The SMGPS is a 1.3 GHz continuum radio survey with $8\arcsec$ angular resolution and a root mean square (RMS) sensitivity of $\sim10$ --$20 \, \mu \text{Jy beam}^{-1}$ \citep{Goedhart:2024}. It covers almost half of the Galactic Plane and is one of the legacy surveys of the SKA precursor telescope MeerKAT, a 64-dish antenna array situated in the Northern Cape region of South Africa. The details of the telescope have been described in \cite{Jonas:2016}, \cite{Camilo:2018}, and \cite{Mauch:2020}.

The SMGPS survey covers two contiguous blocks in Galactic longitude of $251^{\circ} \leq l \leq 358^{\circ}$ and $2^{\circ} \leq l \leq 61^{\circ}$ with each block typically covering a Galactic latitude of $|b| \leq 1.5^{\circ}$. Due to the Galactic Plane warp, the region with Galactic longitude of $251^{\circ} \leq l \leq 300^{\circ}$ had an adjusted Galactic latitude limit of $-2.0^{\circ} < b < 1.0^{\circ}$. The survey coverage area is shown in Figure 1 of \cite{Goedhart:2024}. The SMGPS observations were made between July 21, 2018, and January 14, 2020, using the $L$-band receiver system, covering a frequency range of 856 to 1712 MHz with 4096 channels, and a correlator integration period of 8 seconds. The observations were performed over a series of $\sim10$ hour sessions, cycling between 9 points on a hexagonal grid spaced by $0.494^{\circ}$ to provide uniform sensitivity. The details of the survey, including the initial data products and results, have been reported in \cite{Goedhart:2024}.

\begin{figure}
	\includegraphics[width=\columnwidth]{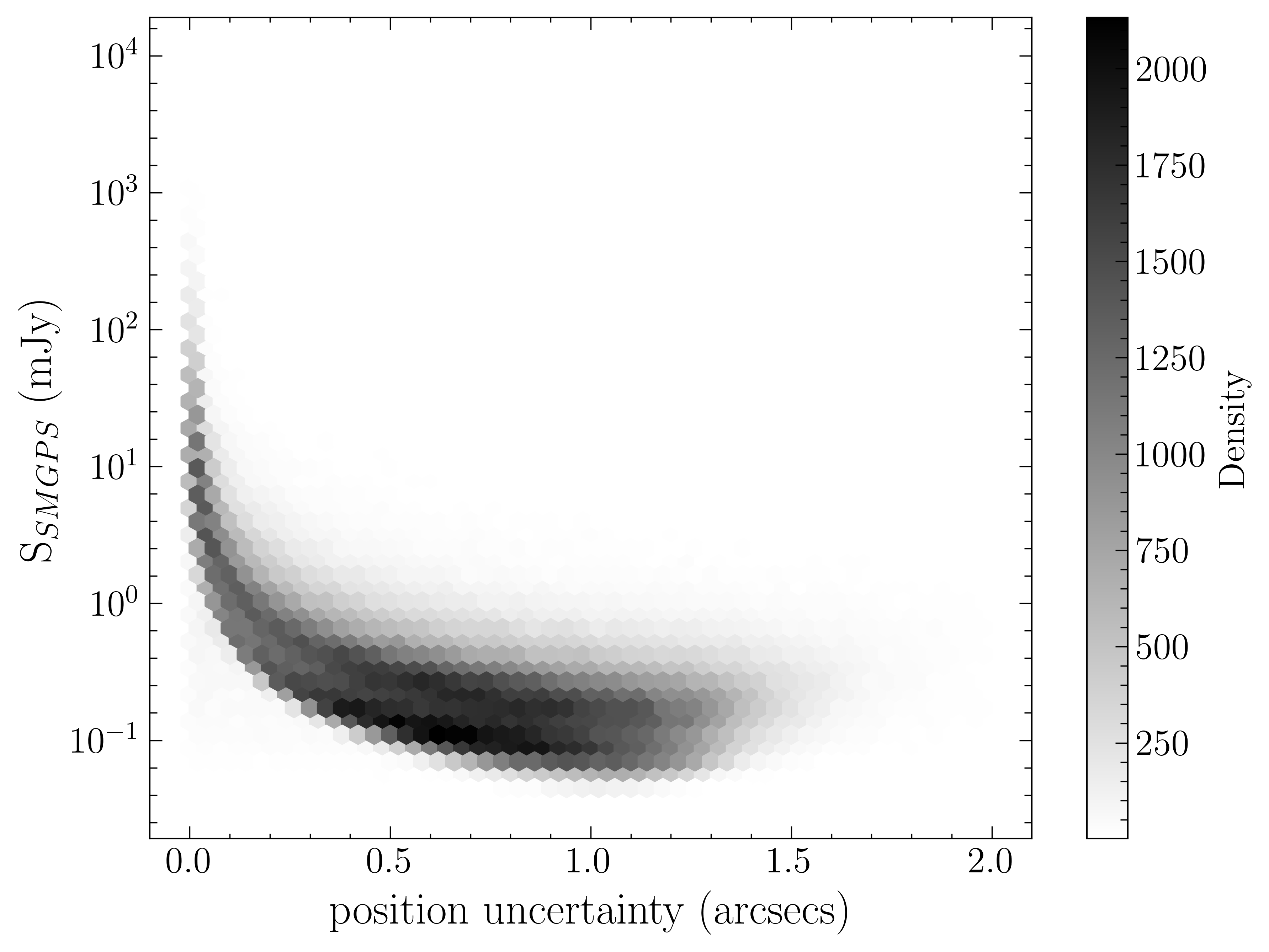}
    \caption{Density plot showing SMGPS flux density (in mJy) as a function of position uncertainty (in arcseconds). The colour intensity indicates regions with a higher concentration of data points.}
    \label{flxpos}
\end{figure}

This paper is based on the SMGPS compact source catalogue (Mutale, et al., {\it in prep.}), which is extracted from the 1.28 GHz zeroth moment images, described in \cite{Goedhart:2024}, resulting in 443,455 unique sources. The majority of the compact sources had positional uncertainties in the range of $0\farcs 1 - 1.5\arcsec$ (see Figure \ref{flxpos}). The compact source positions are in the International Celestial Reference System (ICRS) with a median epoch of J2019.4.

\subsection{Gaia Data Release 3}
\textit{Gaia} Data Release 3 (\textit{Gaia} DR3), which contains $1.81 \times 10^{9}$ sources, is the latest data release catalogue. \citep{GaiaCollaboration:2016a, GaiaCollaboration:2016b, GaiaCollaboration:2023}. It contains astrometric parameters such as position, proper motion, and parallax; and photometric information covering the optical passband (330$-$1050 nm) in the $G$, $G_{\rm BP}$ and $G_{\rm RP}$ filters \citep{Gaia2016, Gaia2022arXiv, hodgkin2021Gaia}. The $G$ passband covers the entire wavelength range. The other two passbands, $G_{\rm BP}$ and $G_{\rm RP}$, cover smaller wavelength ranges, approximately 330 to 680 nm, and 630 to 1050 nm, respectively \citep{Weiler_2018}. 

Beyond the parameters mentioned above, \textit{Gaia} DR3 also provides supplementary catalogues, such as the astrophysical parameters (including effective temperature [T$_{\rm eff}$], surface gravity [$\log g$] and iron abundance [Fe/H]) and the light curves of sources with high variability. 

All \textit{Gaia} source positions are on the ICRS, epoch J2016.0, and for this study have been epoch-propagated to J2019.4 to align with the median ICRS epoch of the SMGPS, using Gaia proper motion information. 

\section{Catalog Cross-matching \& Analysis}

To find SMGPS stellar optical counterparts, several approaches were explored. This included a position cross-match, where we consider all \textit{Gaia} sources within a SMGPS source localization region. In addition, a volume-specific cross-match was applied, where we used \textit{Gaia} sources with specific distance limits. Finally, we applied a stellar population cross-match, where \textit{Gaia} sources belonging to specific stellar population types of known radio emitters were explored. Each of the approaches is described in more detail in the sections below.

\subsection{Full Sample Crossmatch} \label{fsc}
Because the radio sources are in the Galactic Plane, where the \textit{Gaia} object density is high, relying on a basic nearest-source search for \textit{Gaia} optical counterparts will lead to potentially erroneous associations. On the other hand, the angular resolution of MeerKAT can lead to source confusion in crowded fields such as the Galactic Plane. Also, most of SMGPS compact sources are likely to be extra-galactic objects such as galaxies, quasars and active Galactic nuclei (AGN) \citep{Callingham:2019}. To better understand the false association rate between SMGPS and \textit{Gaia} DR3, we performed a Monte Carlo simulation.

The purpose of the simulation is to estimate the rate of spurious matches as a function of the search radius. Our simulation proceeds by randomizing the position of the SMGPS sources. First, we generate a \textit{Gaia} catalogue containing \textit{Gaia} source IDs and coordinates within $120\arcsec$ (2 arcminutes) of each SMGPS source. Then, in our simulations, we randomize SMGPS positions to be within 1 arcmin radius of the actual SMGPS source position in order to assess the probability of false associations.

For each actual SMGPS source, we generate an offset position by generating an offset vector with random position angle (uniformly distributed between 0 and 360 degrees) and a random length between $20\arcsec$ and $60\arcsec$, and add this random vector to the actual position. This creates a ``fake'' radio catalogue sample, which is used to cross-match with \textit{Gaia} and to generate false matches for each iteration. The random length between $20\arcsec$ and $60\arcsec$ ensures that the randomized positions remain within a reasonable distance from the original SMGPS sources while avoiding overlap with (neighbouring) SMGPS true positions. For each iteration, we cross-matched with \textit{Gaia} out to  $5\arcsec$ search radius from the simulated position and calculated the number of matches per radius bin. We performed this iteration 100,000 times and calculated the average number of matches, $N_{\rm MC}$. This provides a statistical background for false positive match frequency.

\begin{figure}
	\includegraphics[width=\columnwidth]{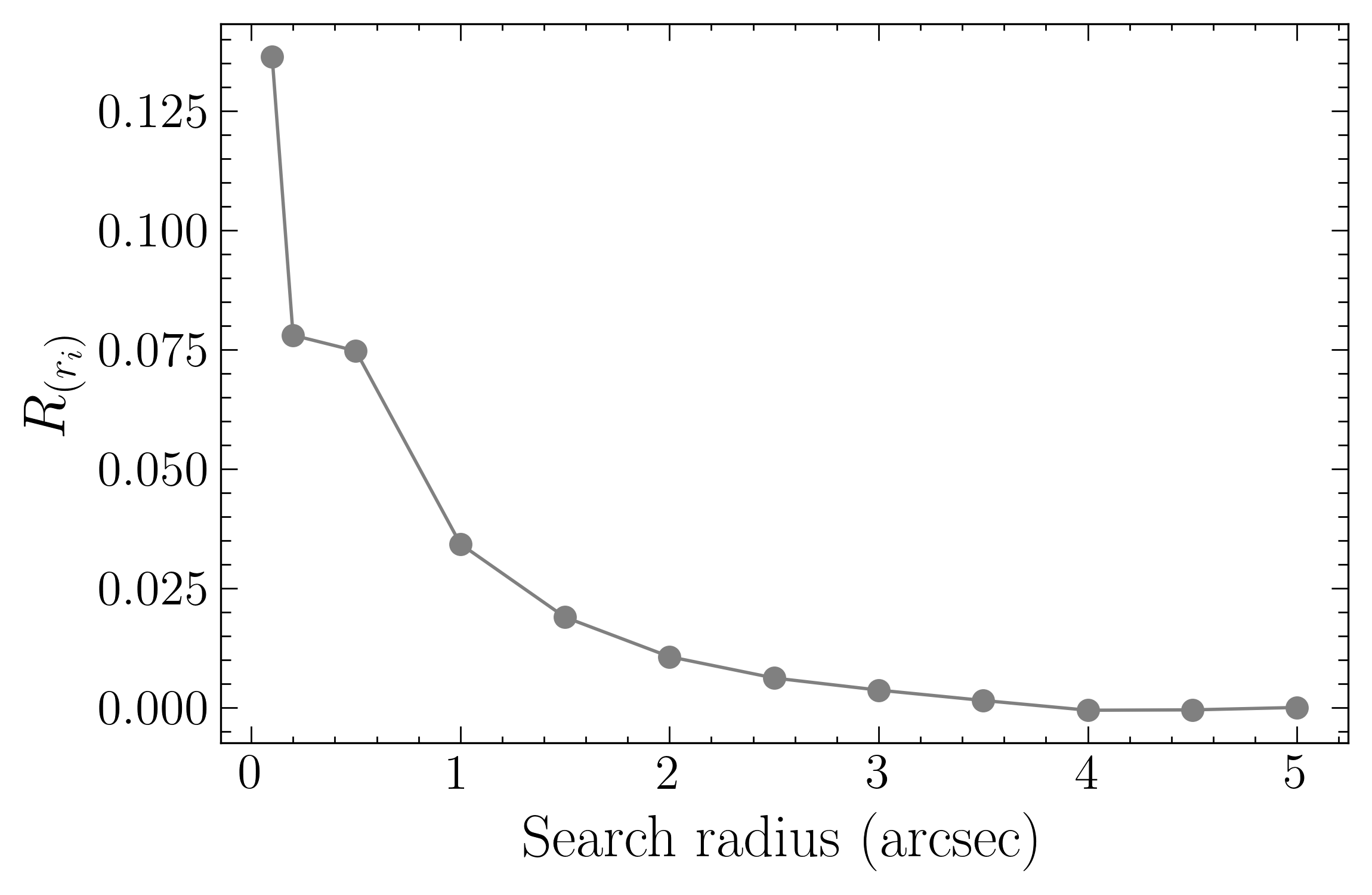}
    \caption{
    A plot of the reliability, $R_{(r_i)}$, against the search radius in arcseconds. 
    } 
    
    \label{xmsim2}
\end{figure}

To assess the reliability of the crossmatch for each search radius bin, we calculate the probability of reliably finding one or more matches within a search radius, using Equation \ref{eqn1},  which defines the reliability $R_{(r_i)}$ as the proportion of SMGPS-\textit{Gaia} matches, reduced by the contribution of random coincidences.
This is similar to the methodology summarized in Section 2.2.3 of \cite{Driessen:2024}.

\begin{equation}
  R_{(r_i)} = 1 - \frac{N_{\text{MC}}(r_i)}{N_{\text{initial}}(r_i)}
  \label{eqn1}
\end{equation}

where $R_{(r_i)}$ is the reliability of the matches in a given search radius $r_i$, $N_{\text{MC}}(r_i)$ is the average number of matches derived from the Monte Carlo simulation, while $N_{\text{initial}}(r_i)$ is the number of matches for the SMGPS cross-match with \textit{Gaia}. The closer $R_{(r_i)}$ is to 1, the higher the reliability of the matches in a given search radius.

In Figure \ref{xmsim2}, we show the reliability, $R_{(r_i)}$ as a function of the search radius $r_i$. The distribution indicates there is a high probability that 1 or more matches in each search radius bin are a chance coincidence and not a true physical match as $R_{(r_i)}$ is always smaller than 0.15. Hence, we conclude that this cross-matching method is not reliable. Therefore, we explored the other approaches in the next sections to reliably find optical counterparts.

\subsection{Volume Specific Crossmatch}
Next, we attempted a volume-specific cross-match based on \textit{Gaia} distance estimates. This is similar to \cite{Yiu:2023}, who searched for LOFAR Two-metre Sky Survey (LoTSS) and VLA Sky Survey (VLASS) stellar sources by cross-matching both radio catalogues with the \textit{Gaia} Catalog of Nearby Stars (GCNS; \citealt{GaiaCollaboration:2021}).

In our case, we started by simulating the random position of all SMGPS sources of SMGPS (as described in the previous section) and cross-matching with \textit{Gaia} samples filtered based on the limits of the distance range of 50, 100, 150, 200, 250, 300, 350, 400, 500, 600, 750, 1000, 1500, 2000, 2500, 3000 and 3500 pc. The \textit{Gaia} distance estimates are from \cite{bailer2021}, who took a Bayesian probabilistic approach in estimating the stellar distances by using a Bayesian prior constructed from a three-dimensional model of the Milky Way Galaxy, which accounts for interstellar extinction and \textit{Gaia}'s variable magnitude limit. This leads to the derivation of two distance estimates, geometric and photogeometric, for the \textit{Gaia} DR3 catalogue sources. In Section \ref{sselect}, we will return to the use of these distances.

\begin{figure}
\centering

\begin{subfigure}{\columnwidth}
    \includegraphics[width=\linewidth]{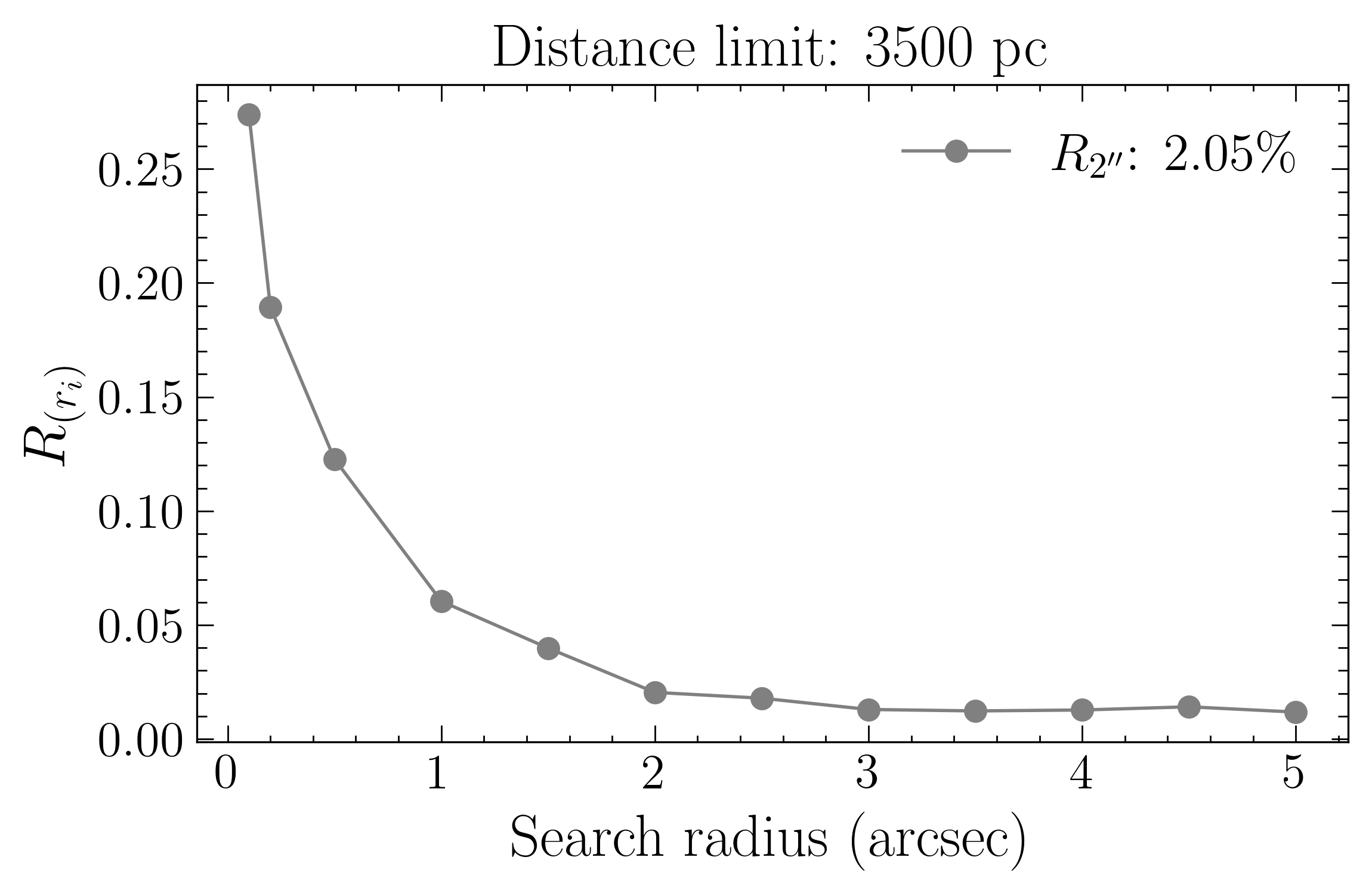}
\end{subfigure}

\begin{subfigure}{\columnwidth}
    \includegraphics[width=\linewidth]{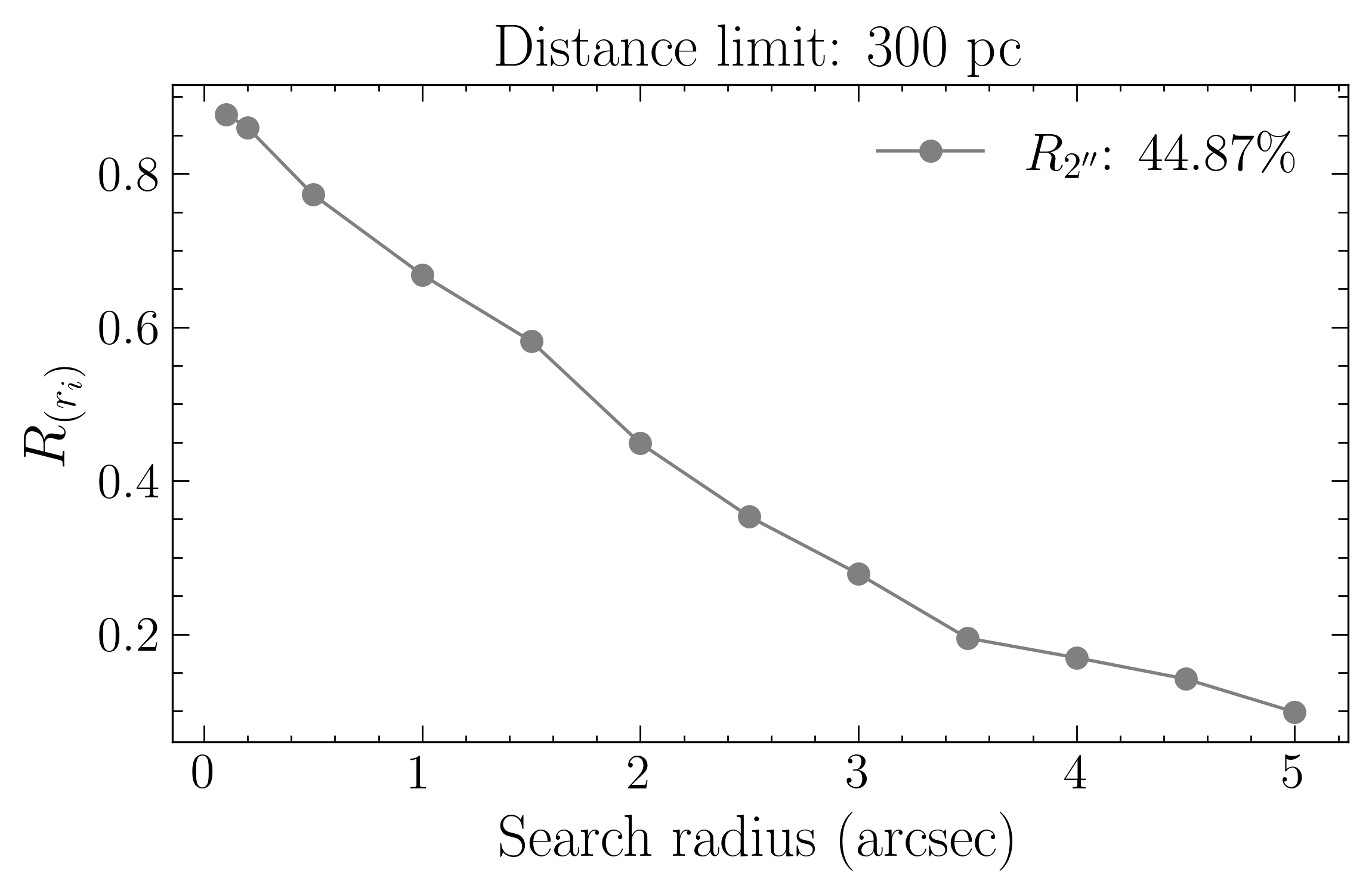}
\end{subfigure}

\begin{subfigure}{\columnwidth}
    \includegraphics[width=\linewidth]{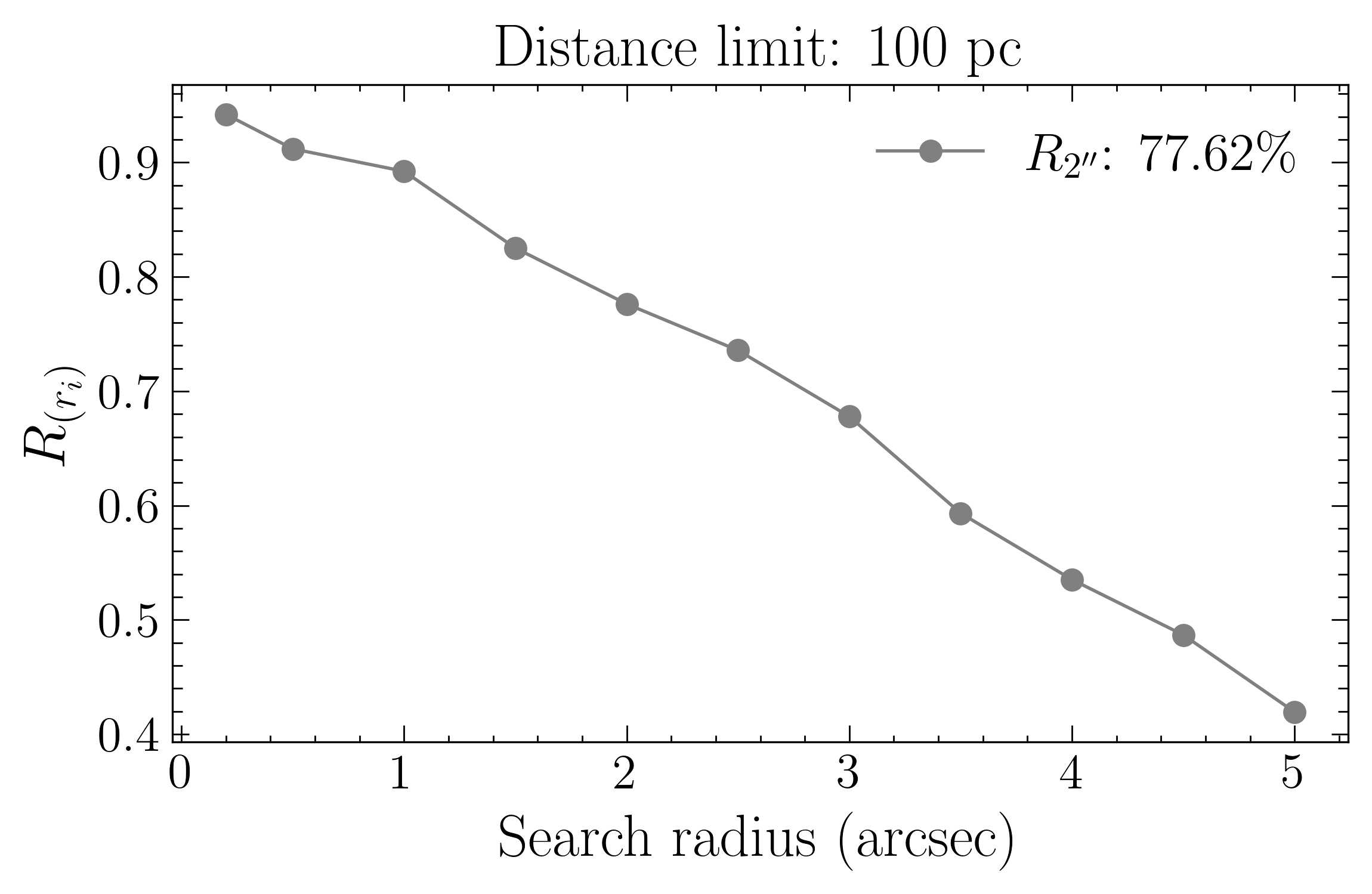}
\end{subfigure}

\begin{subfigure}{\columnwidth}
    \includegraphics[width=\linewidth]{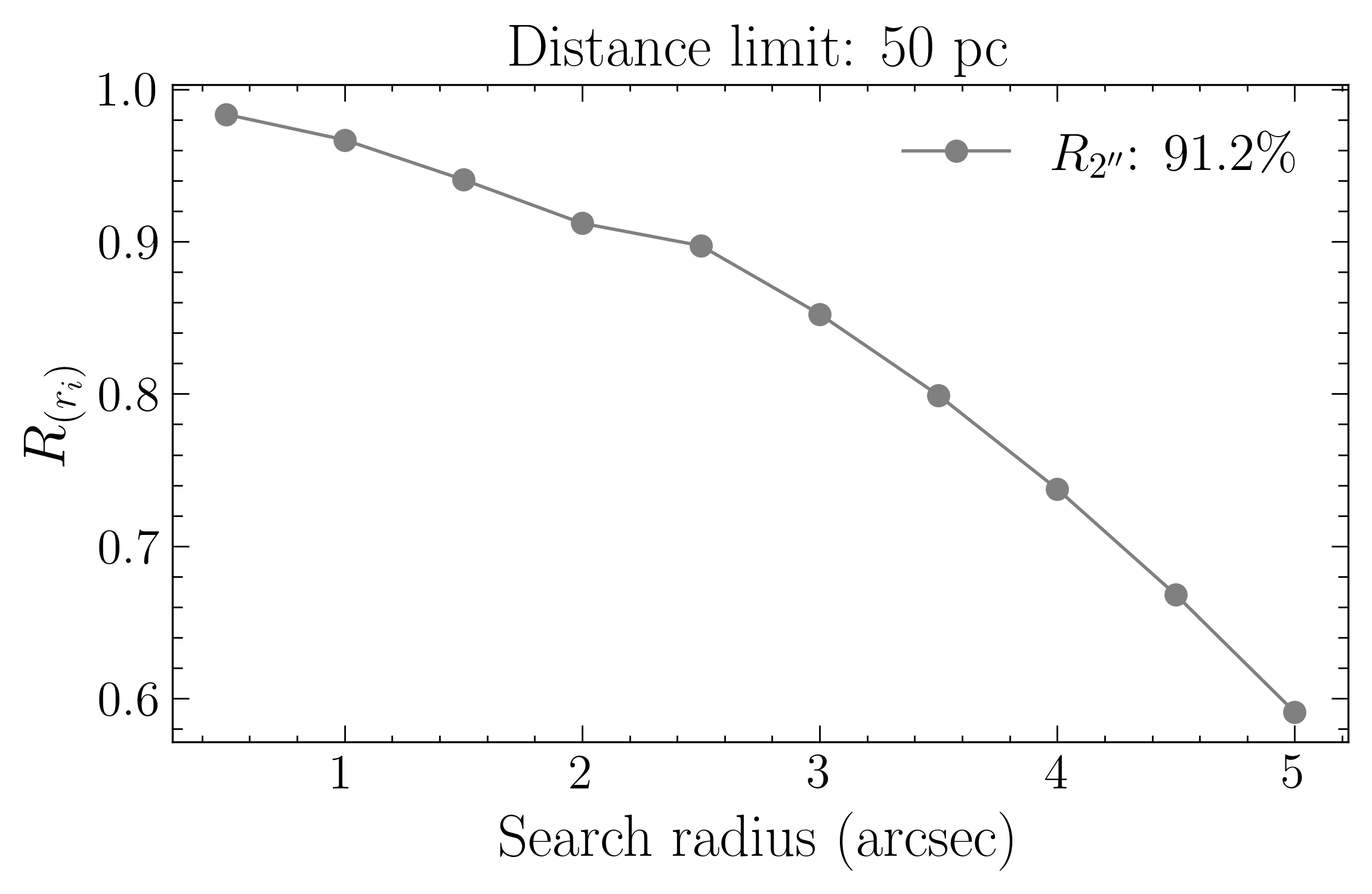}
\end{subfigure}

\caption{Plot showing the reliability, $R_{i}$ out to a search radius of 5\arcsec, for volume-limiting distances of 3500, 300, 100, and 50 pc, respectively. Noted in each plot is the reliability at a specific radius of 2\arcsec, $R_{2\arcsec}$}
\label{distxm}
\end{figure}

We repeated the process outlined in Section \ref{fsc} by cross-matching the SMGPS simulated positions with \textit{Gaia} sources within the above distance limits. This was repeated 100,000 times, and the average number of matches was estimated per search radius. In Figure \ref{distxm}, we show the reliability, $R_{(r_i)}$ per search radius for each distance scale. For space constraints, we only show the distance limits of 3500, 300, 100 and 50 pc. Also noted in the plot is the reliability at $2\arcsec$ cross-match radius, $R_{2\arcsec}$ for all distance limits (see also Table \ref{reliability}). Even at distances of  only a few hundred parsecs, the reliability, $R_{(r_i)}$ is relatively poor at only 20-60\%,  and the number of matches is low. For example, at 100 pc and 50 pc, there are only 25 and 6 stars, respectively, with reliabilities $R_{2\arcsec}$ of 78\% and 91\%. This highlights the limitations of volume-specific cross-matching in the Galactic Plane based on stellar distance, as it yields a small number of sources with reliable matches.

\begin{table}
\centering
\caption{Cross-match reliability at $2\arcsec$ search radius at different distance limits in pc. The number of \textit{Gaia} sources within the $2\arcmin$ footprint of SMGPS is shown in the second column.}
\label{reliability}
\begin{tabular}{l|l|l|l}
\hline
Distance Limit (pc) & Footprint No & $R_{2\arcsec}$  (\%) & Number of matches \\ 
\hline
3500 & 16557624 & 2.05 & 16293 \\
2500 & 8553838 & 2.26 & 8534 \\
1500 & 3045344 & 3.06 & 3130 \\
1000 & 1286522 & 9.11 & 1450 \\
750 & 708733 & 11.93 & 829 \\
500 & 307092 & 22.72 & 406 \\
300 & 98798 & 44.87 & 178 \\
200 & 39522 & 61.74 & 101 \\
100 & 5872 & 77.62 & 25 \\
50 & 554 & 91.20 & 6 \\

\hline
\end{tabular}
\end{table}

\begin{figure*}
	\includegraphics[width=\linewidth]{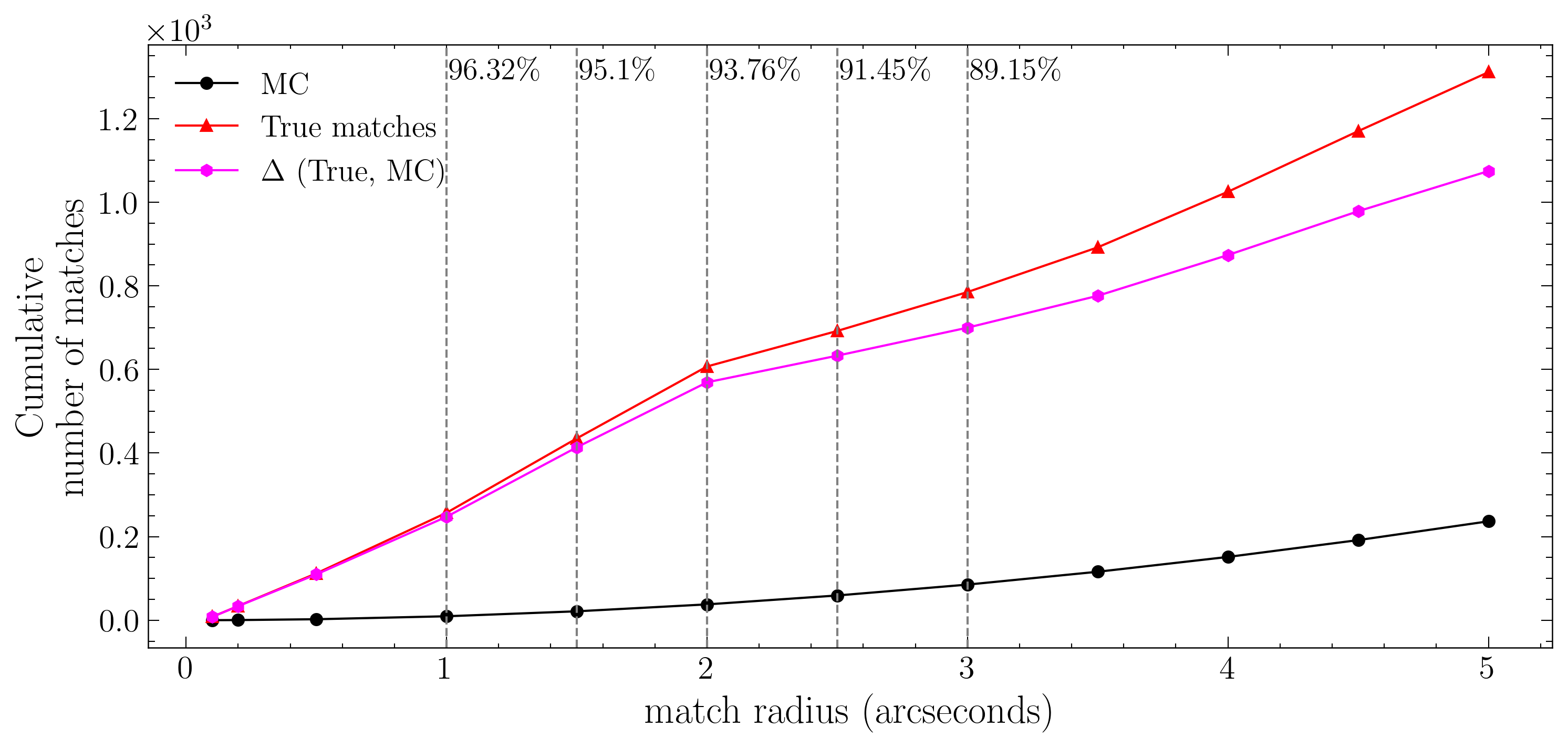}
    \caption{Cumulative distribution showing the number of matches at a specific search radius. The red and black lines represent the number of matches derived from the cross-match between SMGPS positions and the Monte Carlo simulation positions with \textit{Gaia}, respectively. The magenta line represents the difference between the two distributions, which are judged to be true associations. The dashed vertical lines represent different search radii, and the reliabilities, $R_{(r_i)}$ for each search radius, is indicated at the top of the diagram.}
    \label{xmclean}
\end{figure*}


\subsection{Stellar Population Sampling} \label{spsamp}
Given the limitations of the full \textit{Gaia} sample cross-match and the volume-specific cross-match, we explored an alternative approach. Stellar radio emission has been linked to specific stellar populations, including flare stars, main-sequence stars like RS CVn binaries and BY Dra systems, young stellar objects (YSOs), and early-type stars such as OB binaries and Wolf-Rayet stars. 
Furthermore, X-ray-emitting stellar sources are promising candidates for radio emission, particularly coronally active stars \citep{Matthews_2018radio}, which include many of the categories listed above.

A catalogue of these stellar populations was compiled from a review of the literature. \cite{Yang:2023} compiled a catalogue of optical flare events from the TESS survey, while \cite{Rate:2020} assembled a catalogue of Galactic Wolf-Rayet stars. Furthermore, \cite{Melnik:2020} and \cite{Carretero-Castrillo:2023} investigated the Galactic OB stellar sample. We also collected variable star samples from \cite{Heinze:2018} and \cite{Chen:2020}, who studied variable stars from the ATLAS and ZTF surveys, respectively. Furthermore, we used the TESS cool dwarf catalogue by \cite{Muirhead:2018}.

The characteristics of stellar radio source candidates at X-ray wavelengths were derived from observations by the XMM-Newton, ROentgen SATellite (ROSAT), and the extended ROentgen Survey with an Imaging Telescope Array (eROSITA) surveys  \citep{Freund:2018, Freund:2022, Freund:2024}. We combined these catalogues with the previous ones mentioned in the last paragraph, resulting in a sample of 360,959 \textit{Gaia} sources within the SMGPS footprint. Although some catalogues included \textit{Gaia} identifications, others did not. In these cases, we searched the \textit{Gaia} catalogue and retrieved the corresponding Gaia source IDs and positions. Also, all \textit{Gaia} positions were epoch-propagated to J2019.4. This approach is similar to \cite{Driessen:2024}, where similar stellar populations were considered for finding radio stars in the ASKAP Sydney Radio Star catalogue. We also point out that by only selecting these populations, we may miss sources that could still be stellar radio emitters but whose characteristics do not match with known classes.

The cross-match simulation was repeated as in the previous sections. We compared the number of cross-matches obtained from the SMGPS positions and the Monte Carlo simulations, and the cumulative distribution is shown in Figure \ref{xmclean}. In this cross-match case, there is a significant difference between the cross-matches of SMGPS positions and the simulated MC ones, indicating that most of the SMGPS-\textit{Gaia} matches are likely to be physically associated. This is supported by the reliabilities $R_{(r_i)}$ estimated at different search radii. For example, at  2\arcsec, the reliability was estimated at 
94\%.

\subsection{SMGPS - Gaia Positional Chance Alignment Analysis} \label{mmrel}
While the last section has shown an improvement in the number of reliable cross-matches, we took an additional cross-match step. Using a different approach, we tried to assess the reliability of the matches between SMGPS and \textit{Gaia} DR3 on a case-by-case basis. To achieve this, we considered two key metrics: $f_0$, which quantifies the initial positional agreement between the SMGPS source and the nearest \textit{Gaia} source, and
$S_0$, which provides a statistical measure of the likelihood that the observed positional coincidence is due to a physical association rather than a chance alignment.

To compute $f_0$ and $S_0$, we employed the following procedure:

\begin{enumerate}
    \item Calculate $f_0$: This is the normalized initial offset in arcseconds between the SMGPS source and its nearest \textit{Gaia} source, calculated using the formula below, 
    \begin{equation}
        f_0 = \frac{\sqrt{(\text{dx})^2 + (\text{dy})^2}}{\sigma_{\text{radio}}}
        \label{fo}
    \end{equation}
    Here, $\rm dx$ and $\rm dy$ represent the differences in right ascension and declination, respectively, while $\sigma_{\rm radio}$ is the $ 1 \sigma$ positional uncertainty of the SMGPS source. We did not account for the position uncertainties of the \textit{Gaia} source, as they are insignificant compared to those of SMGPS, earlier shown in Figure \ref{flxpos}. It is important to note that for high proper motion stars, the \textit{Gaia} errors on the proper-motion-propagated positional errors may not be negligible due to the uncertainty on the proper motion itself. However, high proper motion stars constitute only a small fraction of the initial \textit{Gaia} sample used in this work (about 1.2\% for sources with proper motion of $\geq 20$ mas/yr), and the impact of any positional discrepancy is expected to be minimal.
    
    \item Simulate new positions: Using the SMGPS source position, we generate $N_{\rm trial}$ random positions by shifting the SMGPS source randomly and thereafter, compute the new normalized offsets $f_i$ for the nearest \textit{Gaia} source for each trial. In this case, we use an $N_{\rm trial}$ of 10,000. Each shift performed on the SMGPS source is within a $30\arcsec$ radius from the original SMGPS position.

    \item Count $f_i \leq f_0$: Next, we count how many of these 10,000 simulated positions have $f_i$ less than or equal to $f_0$.

    \item Calculate $S_0$: Furthermore, we estimate the fraction of simulated trials where $f_i \leq f_0$ using the relation:

    \begin{equation}
        S_0 = \frac{\text{Number of simulated positions with } f_i \leq f_0}{N_{\text{\rm trial}}}
        \label{so}
    \end{equation}

    where $S_0$ represents the chance-alignment level of the match, with lower values indicating higher confidence that the match is real.

\end{enumerate}

\begin{figure}
	\includegraphics[width=\linewidth]{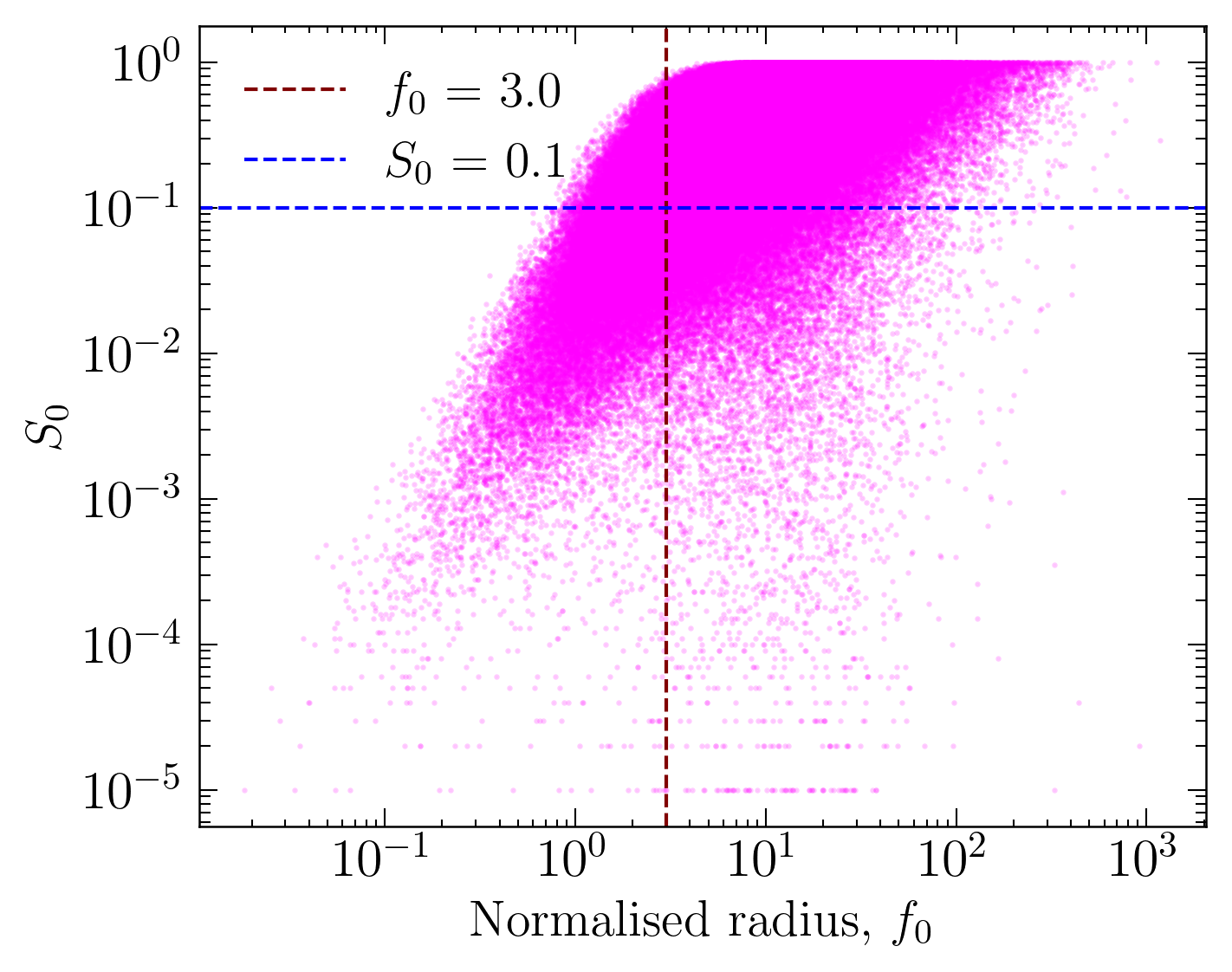}
    \caption{Confidence level, $S_0$ against normalized distance, $f_0$. The dashed lines represent the cutoff for $S_0$ and $f_0$ that were considered for further analysis.}
    \label{fosim}
\end{figure}

The derived normalised radius, $f_0$ and chance-aligment level, $S_0$, from this method is shown in Figure \ref{fosim}. One key takeaway from the diagram is that the majority of the SMGPS sources and \textit{Gaia} are unrelated with poor confidence levels. All sources lying within  $f_0 \leq 3$ and $S_0 \leq 0.1$ were considered for further analysis in the next section.

\subsection{Sample Selection}\label{sselect}
In this section, we consider the selection of SMGPS radio stellar candidates based on the last two subsections  \ref{spsamp} and \ref{mmrel}.

\begin{figure}
	\includegraphics[width=\columnwidth]{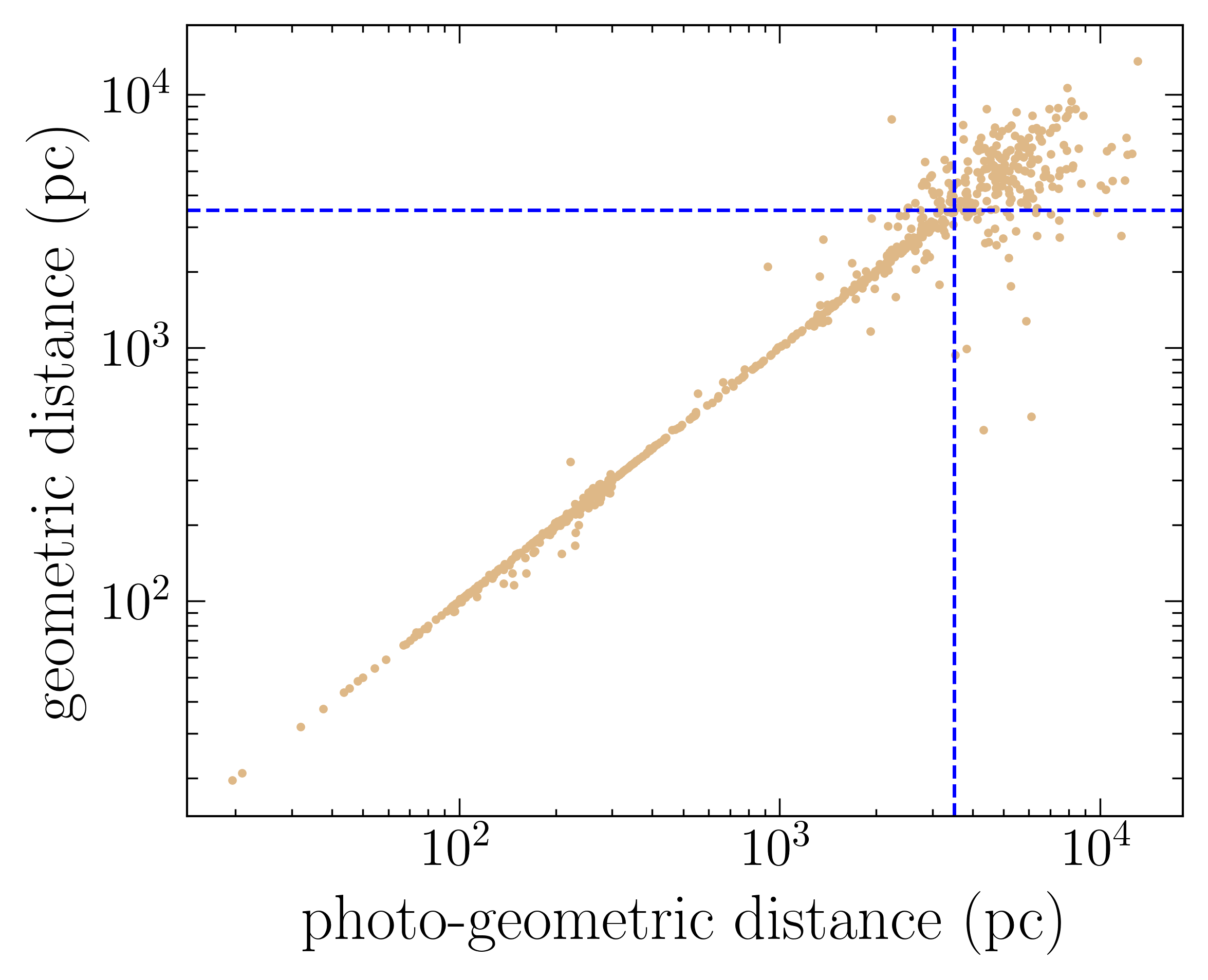}
    \caption{The distribution of the geometric and photogeometric distances for the 832 sources within $3\arcsec$ radius of SMGPS. The dashed line indicates a distance of 3500 pc where both distance estimates diverge.
    }
    \label{distplot}
\end{figure}

\begin{figure}
	\includegraphics[width=\columnwidth]{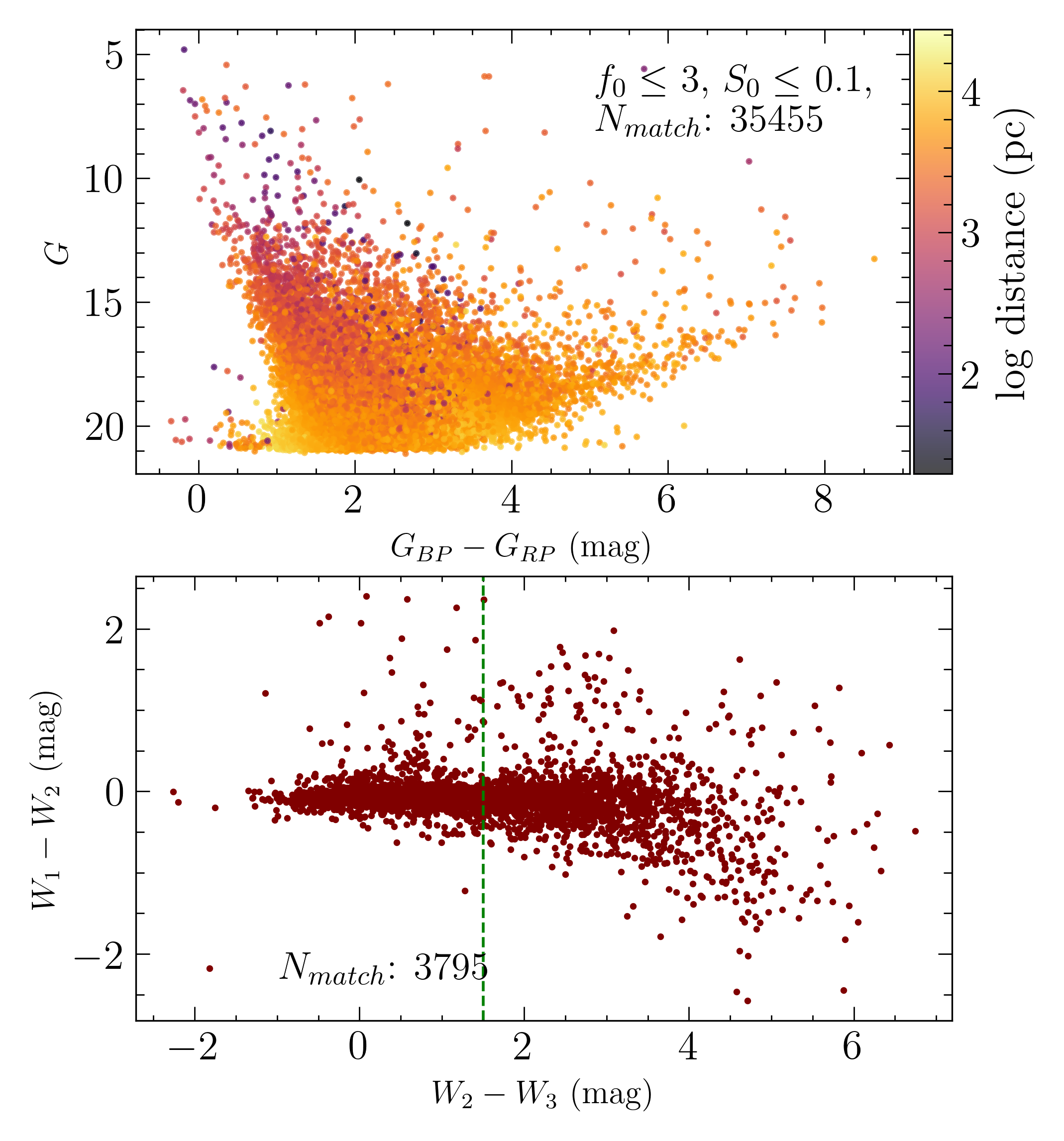}
    \caption{CMD from sources within $f_0 \leq 3$ and $S_0 \leq 0.1$. The top panel shows the CMD from \textit{Gaia}, and the bottom panel is that of AllWISE. The \textit{Gaia} CMD is colour-coded based on the source distance in pc, with a colour bar showing the distance range. The dashed line in the bottom panel represents the AllWISE colour of $W_2 - W_3 = 1.5$, our selection cut-off.}
    \label{fisi}
\end{figure}

\subsubsection{Stellar Population Sample Selection}\label{ssselect}

To select an optical counterpart sample from stellar population sampling (Sect. \ref{spsamp}), we first considered all sources within $\leq 3\arcsec$ of the SMGPS position, which had a cross-match reliability of $\sim 90\%$ or better. To further refine the sample, we compared the distances of the optical sources based on information obtained from \cite{bailer2021}, who derived two distance estimates; the geometric and photogeometric distances, where, geometric, uses the parallaxes and their uncertainties to estimate the distance, whereas photogeometric, combines the parallax, colour and magnitude of the star to estimate the distance. The distribution of both distances is shown in Figure \ref{distplot}, which shows a strong correlation. However, a noticeable scatter is witnessed above 3500 pc. Hence, only sources with geometric and photogeometric distances $\leq 3500$ pc were selected, reducing the sample to 551 stars.  It is important to note that photogeometric distance may implicitly assume single stars rather than binary systems, and may hence provide more reliable distances for single stars.

\subsubsection{Selection based on AllWISE color}\label{rsim}
From the cross-match sample obtained from subsection \ref{mmrel}, we considered sources with $f_0 \leq 3$ and $S_0\leq 0.1$, resulting in 35,455 sources shown in the top panel of Figure \ref{fisi}. The \textit{Gaia} colour-magnitude diagram (CMD) of this sample is shown in the figure with the colour bar reflecting the distances in parsecs. Also shown in the figure is the CMD of AllWISE associations with \textit{Gaia}. One obvious takeaway from the plot is the fact that the majority of the associations are distant and faint, suggesting that they may not be true radio stellar sources, and the radio emission could be from distant extra-galactic sources in the background of optical \textit{Gaia} stars.

To reduce contamination, we looked for AllWISE 
counterparts to our selection of \textit{Gaia} cross-matches, which is one of the surveys available in the external catalogues matched with \textit{Gaia} DR3 \citep{Marrese:2019, Marrese:2022}. AllWISE is an infrared survey that is very sensitive to extra-galactic sources \citep{Wright_2010, Mainzer:2011}. Therefore, it can be used to filter out extragalactic contamination in our sample.

For regions within the Galactic Legacy Infrared Midplane Survey Extraordinaire (GLIMPSE), we also found sources with GLIMPSE emission \citep{Benjamin:2003}. The GLIMPSE (Galactic Legacy Infrared Mid-Plane Survey Extraordinaire) is an infrared survey conducted by the Spitzer Space Telescope, providing high-resolution images of the Galactic plane in the $3.6\,\mu\text{m}$ and $4.5\,\mu\text{m}$ bands \citep{Benjamin:2003}. It is particularly useful for studying stars and star-forming regions, as it detects emission from cooler stars and regions obscured by dust in optical wavelengths. However, as there is no full overlap between GLIMPSE and the SMGPS, we used the AllWISE source list for a full cross-match.

When using AllWISE for source characterization, the colour-colour, $W_1 - W_2$ versus $W_2 - W_3$ phase space diagram is ideal for separating Galactic and extra-galactic sources, see Figure 12 of  \cite{Wright_2010} for more details. For example, \cite{Chen:2018} used it as part of the method to identify AllWISE variable stars. In our case, we selected AllWISE counterparts to \textit{Gaia} with AllWISE $(W_2 - W_3)<1.5$ mag. Since AllWISE position uncertainties are $<1\arcsec$ in most cases, we also restricted our selection to sources with \textit{Gaia} and AllWISE separations within $1.0 \arcsec$. Based on the top panel of Figure \ref{fisi}, the majority of the sources are faint and distant. Hence, we also limited our sample to sources within 1.5 kpc, resulting in 141 AllWISE stars. It is important to note that the AllWISE colour selection will result in the selection of mostly red stars.

\section{Final Sample and Discussion}


Comparing 551 stars and 141 stars selected from the \ref{ssselect} and \ref{rsim} subsections, there are 63 stars common with both samples, so combining both selections results in 629 unique stars which are strong candidates for optical counterparts to SMGPS sources. While this catalogue presents a valuable resource for further follow-up observations, we acknowledge that it lacks a comprehensive assessment of completeness and level of potential contamination. Completeness, particularly for matching rather than detection, is challenging to assess and may vary across different regions of the survey due to varying noise levels and Galactic diffuse emission.

\begin{figure}
	\includegraphics[width=\columnwidth]{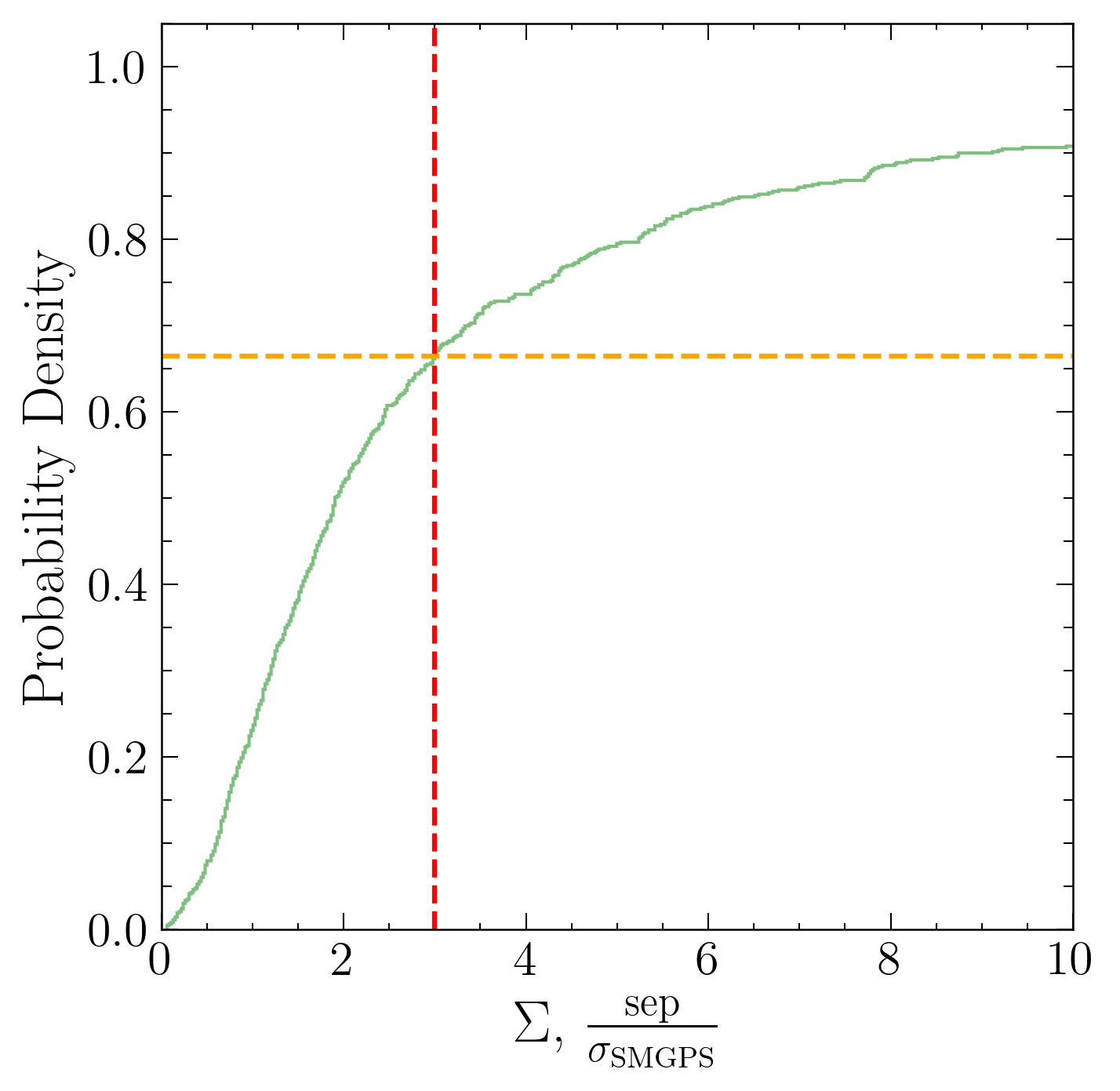}
    \caption{A cumulative distribution of $\sigma = \frac{\text{sep}}{\sigma_{\rm SMGPS}}$ of the stellar candidates with 66 \% of the stellar candidates within $3\sigma$.}
    \label{sigg}
\end{figure}

We assessed the parameter $\Sigma = \frac{\text{\it sep}}{\sigma_{\rm SMGPS}}$, for the matched stellar candidates, where {\it sep} denotes the positional separation between the matched optical and radio sources and $\sigma_{\rm SMGPS}$ is the uncertainty on the radio position. For a fully and properly matched sample we expect a Gaussian distribution with a width of $\sigma_{\rm SMGPS}$. The distribution of $\Sigma$ (shown in Figure \ref{sigg}) indicates that $\sim$70\% of the candidates fall within the $3 \sigma$ radio uncertainty radius. The wider than expected distribution can be due to $i)$ spurious matches, despite our best efforts, $ii)$ our assumption that the optical positional uncertainty is neglible, including caused by the positional uncertainty introduced by the forward propagation based on the proper motions and the uncertainties thereon, but also the fact that some of the radio source positional uncertainties are extremely small ($<<1\arcsec)$ and therefore may not dominate the positional association, or $iii)$ a systematic uncertainty due an astrometric reference frame mismatch between the radio positions and the optical positions: even though both are on the ICRS, it is hard to assess the accuracy of the astrometric frame calibration of the SMGPS sample as outlined in \citep{Goedhart:2024}.

\subsection{Extinction corrected Colour - Magnitude Diagram}
To understand the intrinsic properties of Galactic stellar sources, we estimate the visual extinction $A_V$ of the sources using the dust extinction map from \cite{Lallement:2019} in combination with \cite{Marshall:2006} in the inner disc of the Galaxy. A smooth transition between both models has been established to ensure consistent values. The code computes $A_V$  using the Galactic coordinates ($l$, $b$) and distance estimates from \cite{bailer2021}. With the extinction parameter, the \textit{Gaia} $G$, $G_{BP}$ and $G_{RP}$ mags were corrected for extinction and intrinsic colours were calculated. Additionally, the absolute mag, $M_G$, of the sources was calculated using the extinction corrected $G$ mag and the distance estimates obtained from \cite{bailer2021}

\begin{figure*}
	\includegraphics[width=\textwidth]{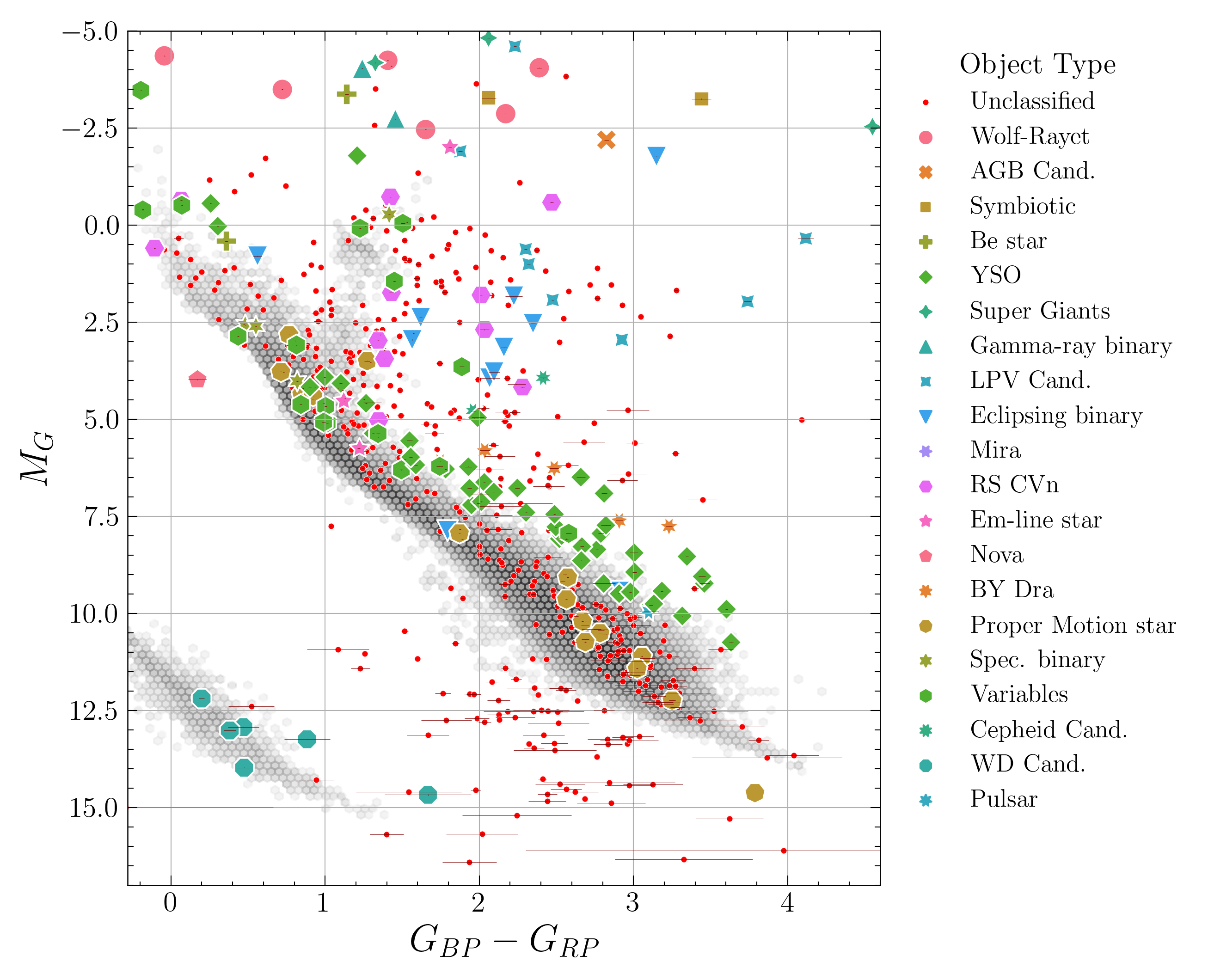}
    \caption{A colour-magnitude diagram showing the intrinsic colour and the absolute magnitude of the 629 SMGPS-\textit{Gaia} counterparts. The stellar type classification is shown in the legend. The grey circles are all \textit{Gaia} sources within 100 pc of Earth binned in uniform colour and magnitude. The coloured markers represent the 629 SMGPS-\textit{Gaia} matches. The larger markers are the 169 stars already classified in the SIMBAD database. The red dots are unclassified stars with either no information on SIMBAD or have been reported in SIMBAD database as ``Star''.}
    \label{hrdiag}
\end{figure*}

Having derived the intrinsic colour and absolute magnitude information, we plot the CMD of the \textit{Gaia} counterparts in Figure \ref{hrdiag}. The CMD shows diverse stellar sources at different evolutionary stages. A search was performed on SIMBAD for previously reported objects in the sample, and various object types were recovered. Some of the object types found include variable stars, long-period variables, Wolf-Rayet stars, supergiants, RS Canum Venaticorum binaries (RS CVns), and young stellar objects (YSOs), many known to be coronally active stars expected to produce radio emission, and overlapping with our catalog classes used in Sect. \ref{spsamp}. On the other hand, for the massive stars such as Wolf-Rayet types, OB stars, Gamma-ray binaries and supergiants, the dominant mechanism responsible for radio emission can be thermal or non-thermal emission from dense, ionized stellar winds.

\begin{figure}
	\includegraphics[width=\columnwidth]{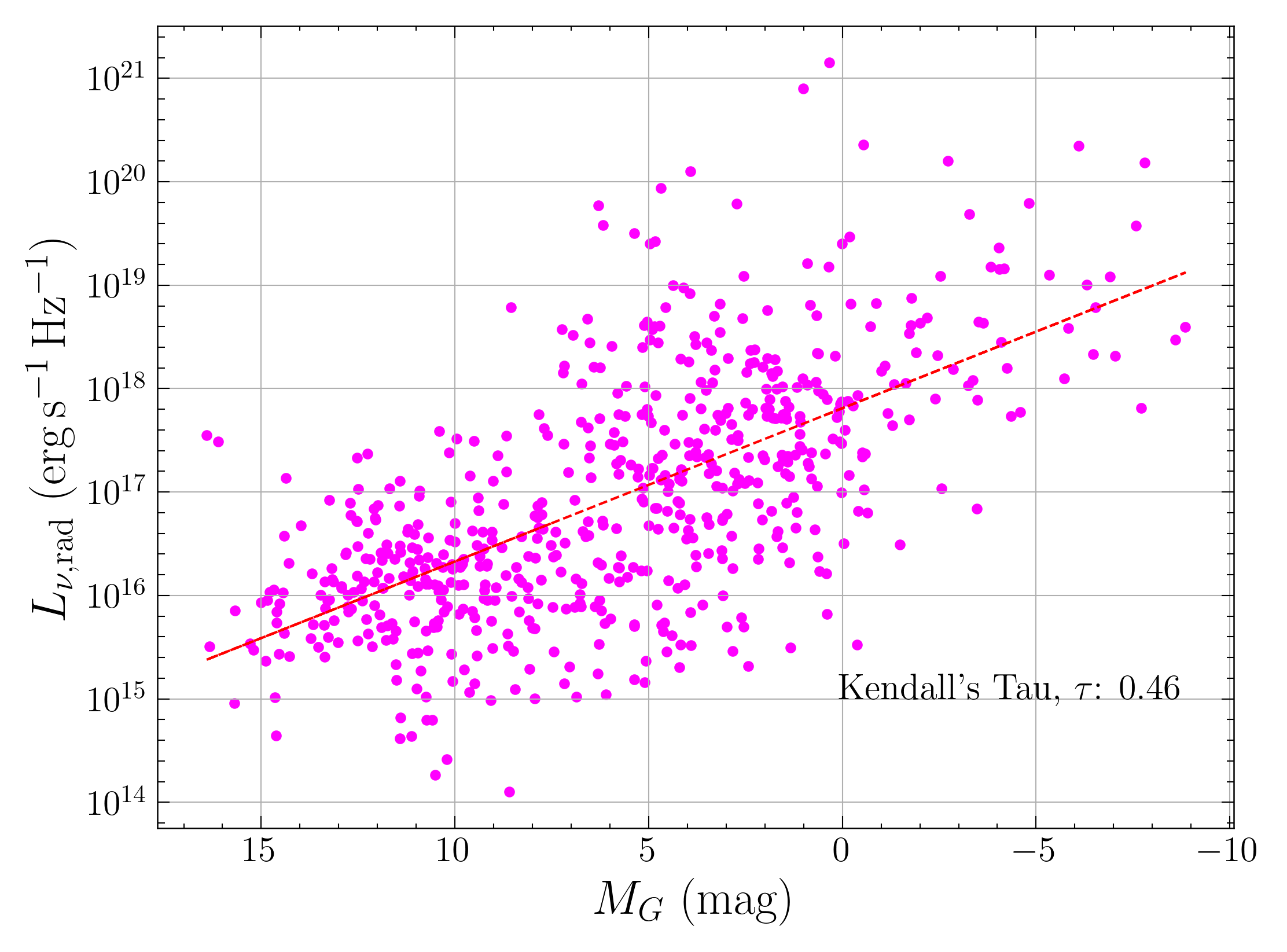}
    \caption{Specific radio luminosity vs. absolute magnitude for the SMGPS-Gaia counterparts. The dashed red line is the linear fit to the two parameters with a Kendall Tau, $\tau$, correlation coefficient of 0.46.}
    \label{abmagrl}
\end{figure}

We further calculated the specific radio luminosities using the relation $L_{\nu} = 4\pi d^{2} S_{\nu}$, where $d$ is the distance and $S_{\nu}$ is the measured SMGPS flux density at 1.3 GHz. The radio luminosity as a function of absolute magnitude is shown in Figure \ref{abmagrl}. A strong correlation exists between the two parameters with Kendall's $\tau$ \citep{Kendall:1938} correlation of 0.46.

\subsection{Source Populations}

The CMD in Figure\ \ref{hrdiag} shows that the MeerKAT-detected sources tend to lie off the Main Sequence, and in areas of either very young stars (YSOs), or massive (upper left) and/or evolved stars (upper right). What these areas in the CMD have in common is the presence of circumstellar gas, either from an accretion disk inflow/outflow, or from a stellar wind or mass ejections (in massive/evolved stars and gamma-ray binaries) or from chromospherically active stars in magnetic binary systems. 

Figure \ref{hrdiag} also shows a few isolated, ``one-off'' objects, such as the old nova V603 Aql, classified as a novalike cataclysmic variable, located below the blue end of the main sequence (independently detected by MeerKAT in the ThunderKAT survey by \citealt{Hewitt:2020}), and two objects between the red part of the main sequence and the white dwarf sequence at bottom-left. We describe each group in more detail. 

\subsubsection{Massive/Luminous Stars}
At the highest luminosities in Figure \ \ref{hrdiag} ($M_G \lesssim -2)$ our associations show a mix of known sources. Prominent amongst these are a number of known Wolf-Rayet systems, Gamma-ray binaries, symbiotic and OB-type binaries. 

We found some massive stars from the SMGPS-\textit{Gaia} associations. The massive stars are mainly OB-type binaries and Wolf-Rayet systems. The OB stars recovered from the \cite{Melnik:2020} OB stellar catalogue are listed in Table \ref{massive-stars}. The Wolf-Rayet stars detected in this sample are listed in Table \ref{wolfrayet}. Of the 9 Wolf-Rayet stars, 8 have been previously detected with either the VLA or the ATCA.

\begin{figure}
	\includegraphics[width=\columnwidth]{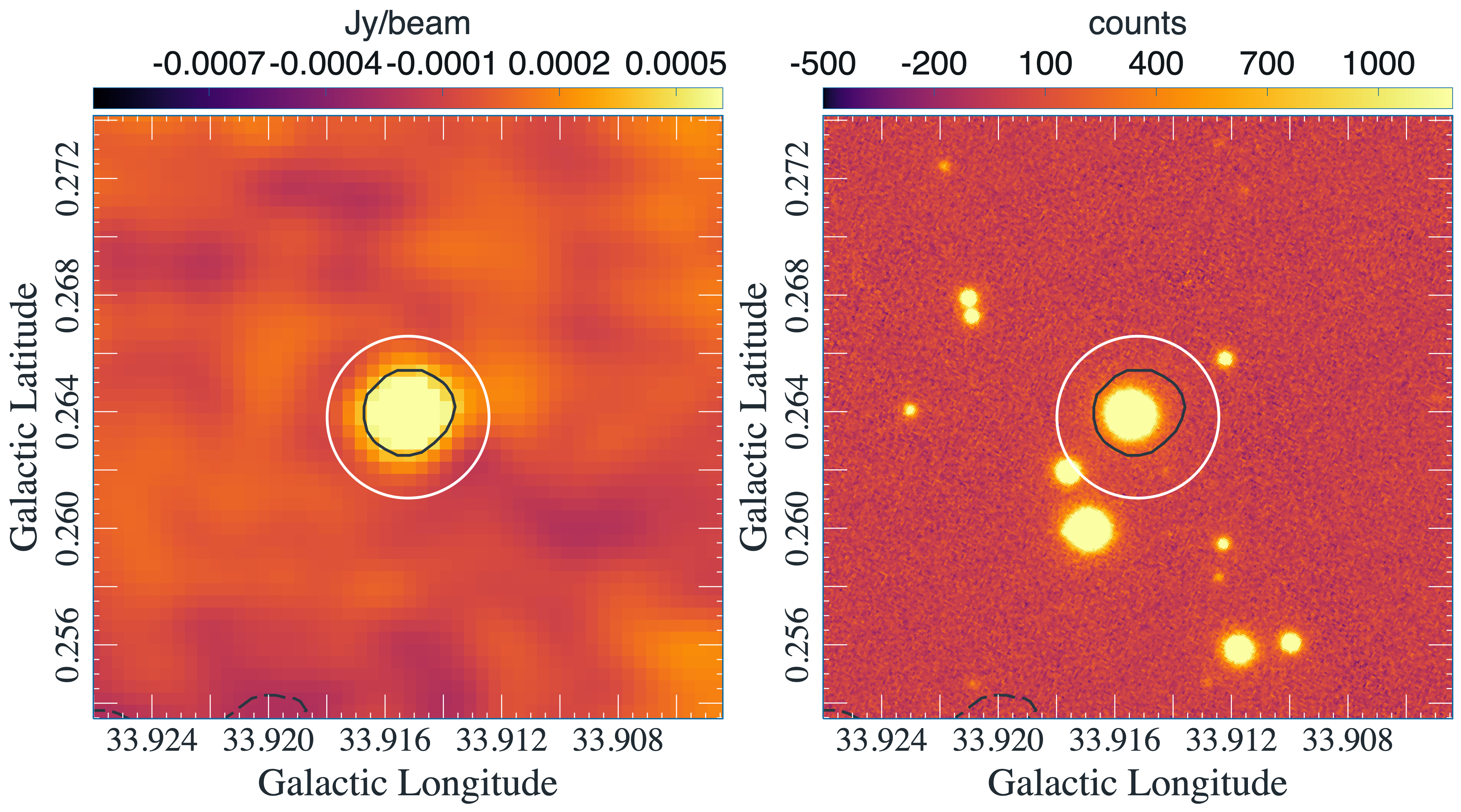}
    \caption{Image view of NaSt1 (WR 122). On the left is the 1.3 GHz SMGPS continuum image. On the right is the PanSTARRS-1 $g-$band image. SMGPS contour is shown in black, and the white circle with \textbf{$10 \arcsec$} radius is centred at the position of the star.}
    \label{ls005}
\end{figure}

\begin{figure}
\centering

\begin{subfigure}{\columnwidth}
    \includegraphics[width=\linewidth]{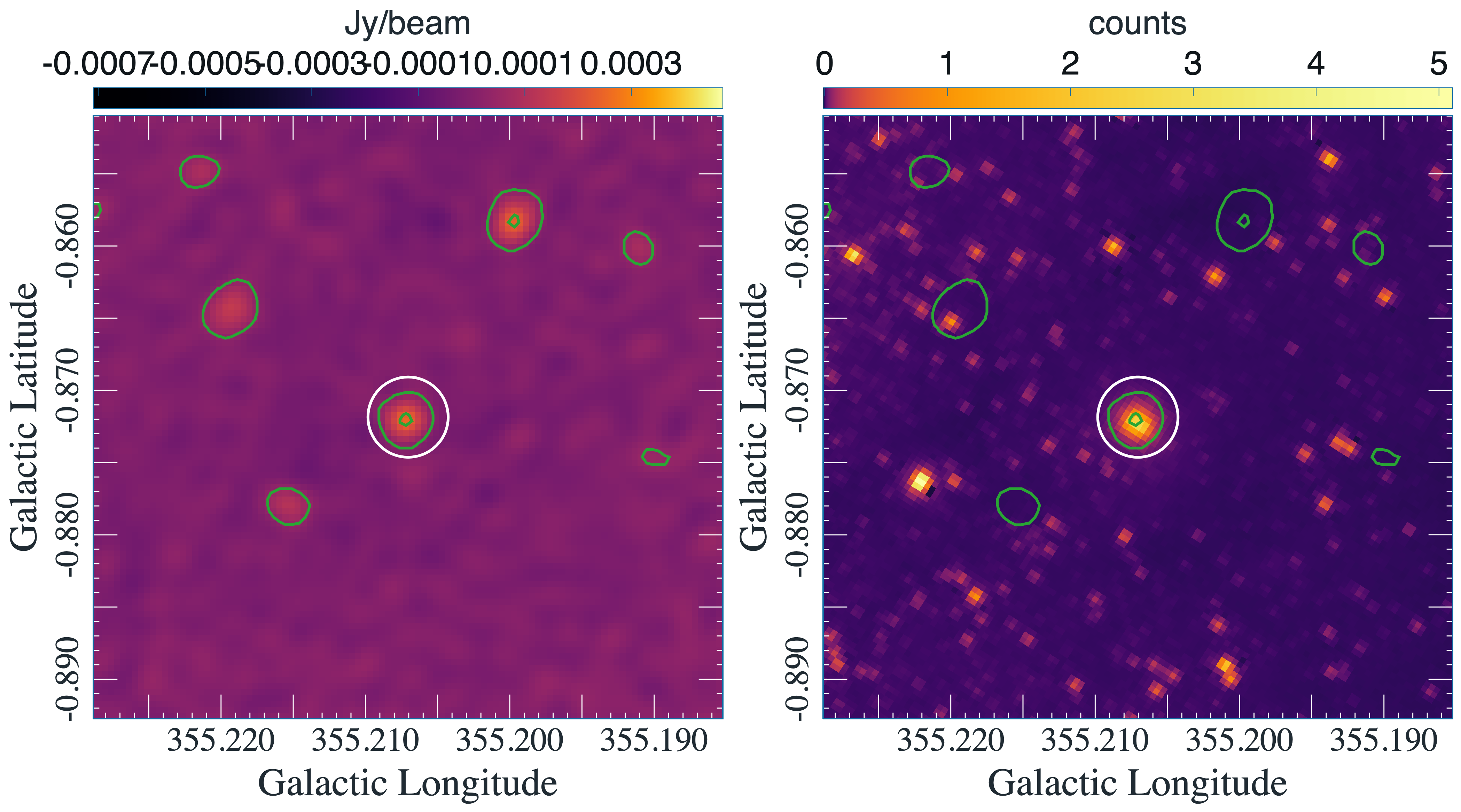}
\end{subfigure}

\begin{subfigure}{\columnwidth}
    \includegraphics[width=\linewidth]{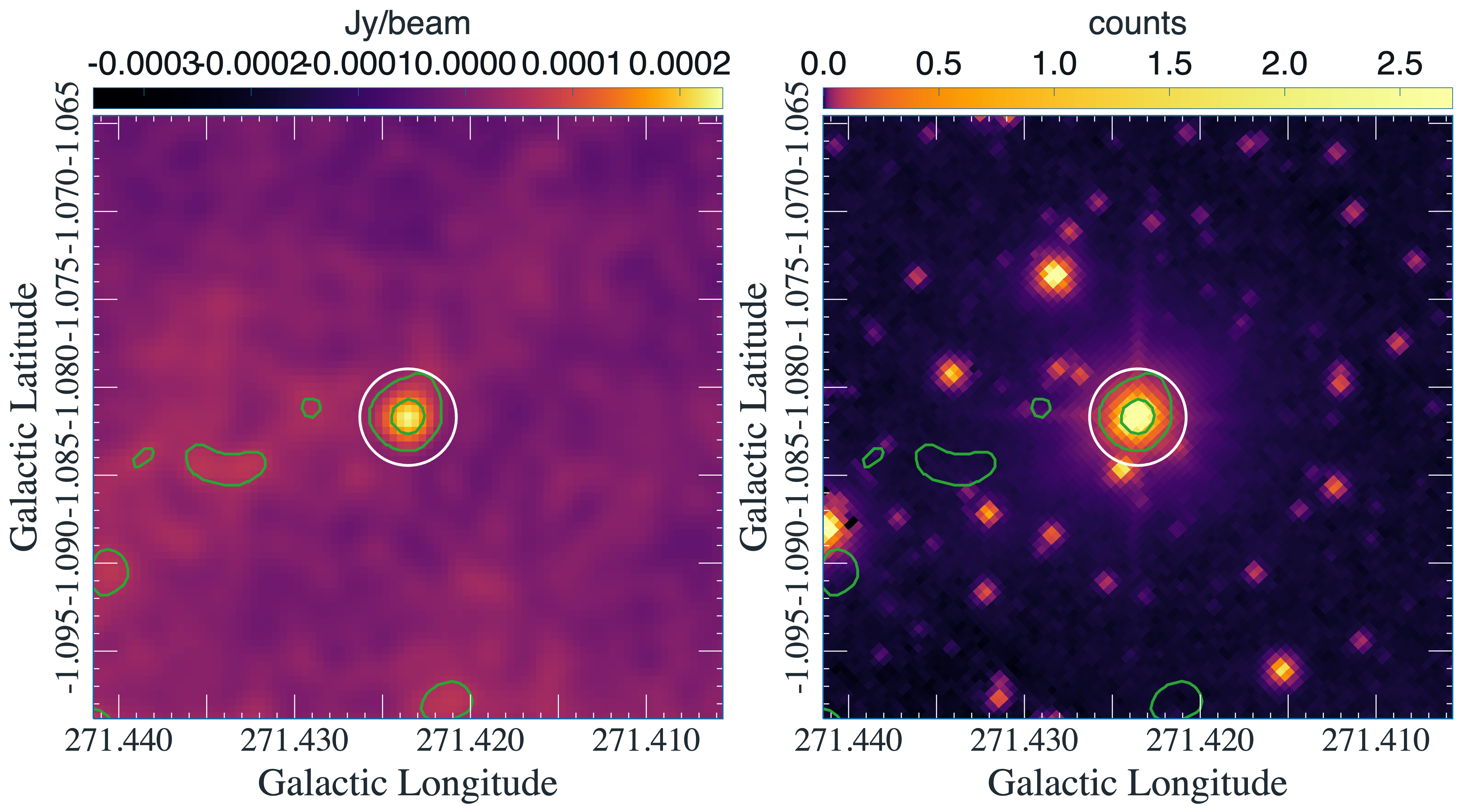}
\end{subfigure}

\begin{subfigure}{\columnwidth}
    \includegraphics[width=\linewidth]{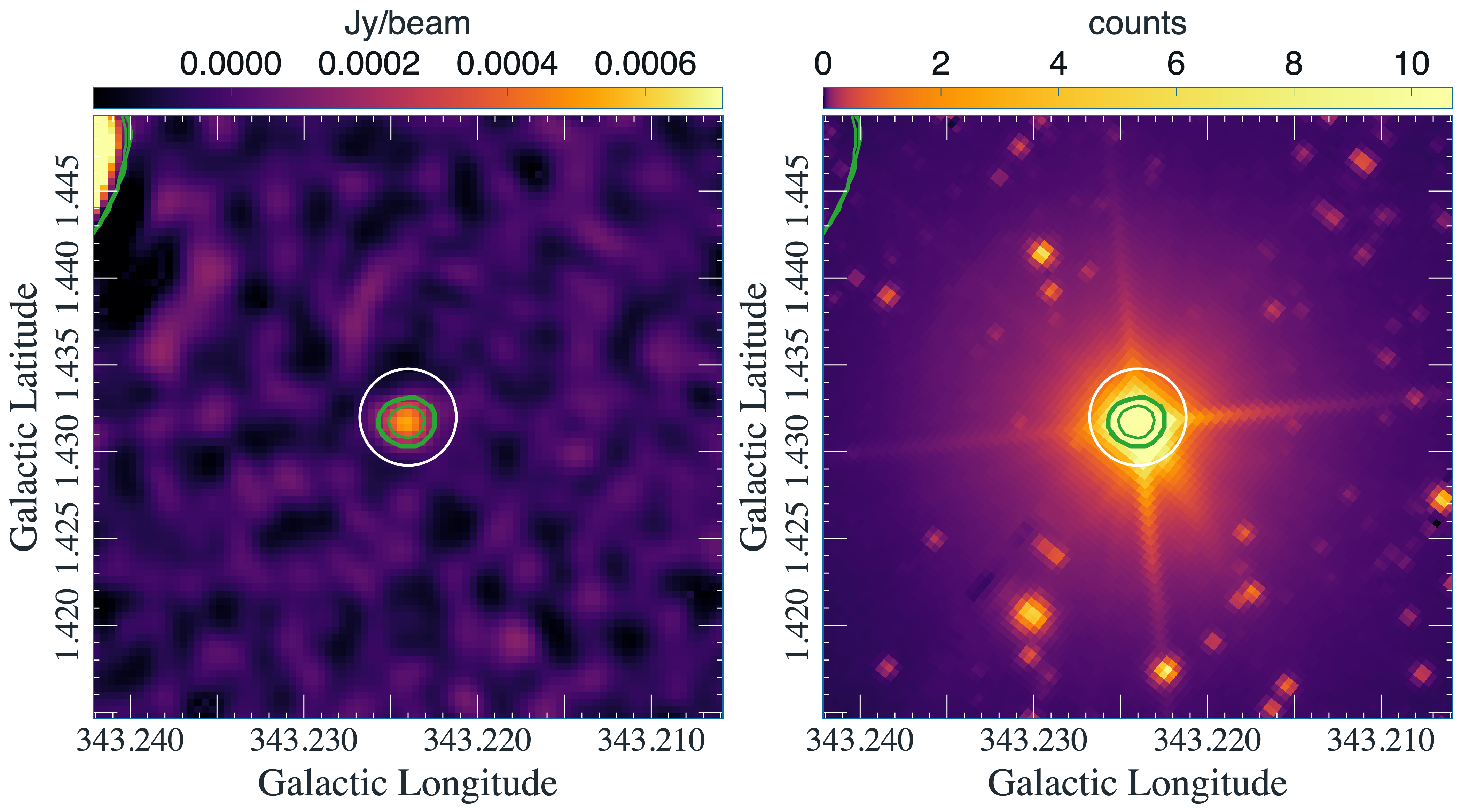}
\end{subfigure}

\begin{subfigure}{\columnwidth}
    \includegraphics[width=\linewidth]{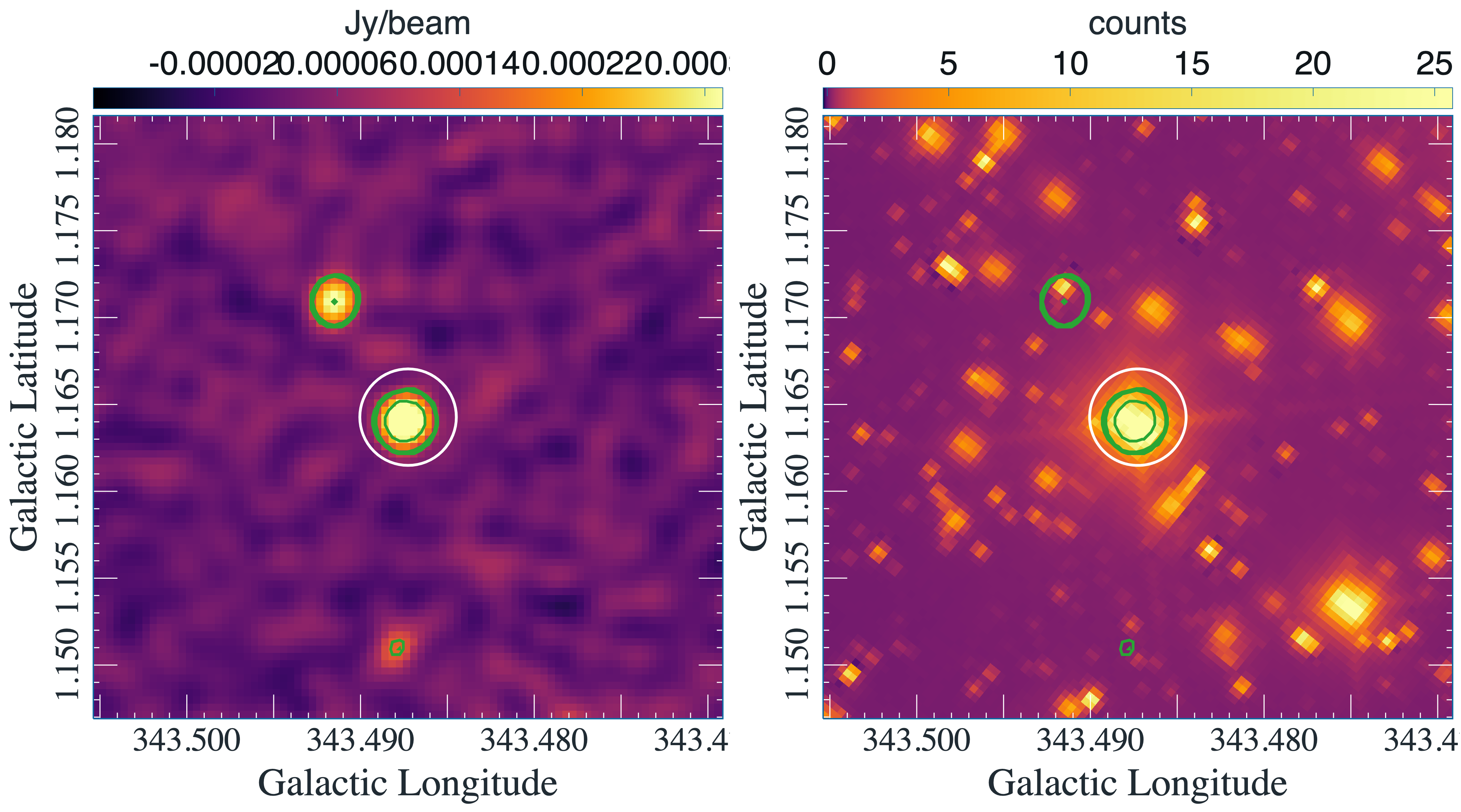}
\end{subfigure}

\caption{MeerKAT and the Dark Energy Camera Plane Survey (DECaPs) view of Wolf-Rayet stars, HD 318016, HD 79573, HD 151932 and HD 1552270. The MeerKAT contour map is overplotted in both images using logarithmic intervals from  $10^{-5} - 10^{1.5}$ Jy. A \textbf{$10 \arcsec$} radius circle is centred at the position of the star in both SMGPS and DECaPs.
}
\label{wolfry}
\end{figure}

We detected radio emission in the region of NaSt1 (WR 122), a late WN10 WR star with strong N II and N III lines \citep{Massey:1983}. The star has an SMGPS flux density of 1.83 mJy. \cite{Mauerhan:2015} reported it as a peculiar emission-line star embedded in an extended nebula of N II emission, with a compact dusty core. The SMGPS and optical PanSTARRS image of NaSt1 is shown in Figure \ref{ls005}. The SMGPS and DECaPS images of four of the previously radio-detected Wolf-Rayet stars from this sample are shown in Figure \ref{wolfry}. \textit{Gaia}-based distances in Tables \ref{massive-stars} \& \ref{wolfrayet} show that the detection limit of the SMGPS allows them to be detected out to at least 3.5 kpc. Given the 3.5 kpc limit we put on any association, this effectively means that the SMGPS is sensitive to an intrinsic radio luminosity in excess of $\sim$10$^{18}$ ergs$^{-1}$ Hz$^{-1}$ for sources within 2 kpc distance.

\begin{table*}
    \centering
    \caption{Massive bright stars from the OB Star catalogue}
    \begin{tabular}{llclcrrll}
    \toprule
    \parbox{1.3cm}{Target Name} & 
    SMGPS ID & 
    \parbox{0.7cm}{ $\sigma_{\rm SMGPS}$ \\ ($\arcsec$)} & \parbox{1cm}{ $S_{1.3\text{GHz}}$ \\ (mJy) } & 
    \parbox{1.4cm}{$L_R$ \\ $(\text{erg}\,\mathrm{s^{-1}}\,\mathrm{Hz^{-1}}
    )$} & 
    \parbox{0.7cm}{$G$ \\ (mag)} & 
    \parbox{0.8cm}{Distance \\ (pc)} & 
    \parbox{0.8cm}{Spectral \\ Type} & 
    \parbox{1cm}{sep. \\ ($\arcsec$)} \\
    \midrule
CD --33 12241 & G355.0639--0.7011 & 0.03 & 16.41 & $1.54 \times 10^{20}$ & 6.49 & 2796.0 & M0 & 1.48 \\
HD 96670 & G290.1971+0.3968 & 0.86 & 0.94 & $1.21 \times 10^{19}$ & 7.34 & 3281.0 & O8. & 0.66 \\
HD 101131 & G294.7783--1.6230 & 0.74 & 1.19 & $1.01 \times 10^{19}$ & 7.07 & 2669.0 & O6. & 0.20 \\
HD 101190 & G294.7814--1.4903 & 0.43 & 0.56 & $3.84 \times 10^{18}$ & 7.26 & 2386.0 & O6. & 0.58 \\
HD 143183 & G328.5405--1.0112 & 0.04 & 5.30 & $3.76 \times 10^{19}$ & 5.83 & 2436.0 & M3 & 0.35 \\
HD 168625 & G014.9778--0.9555 & 0.44 & 24.83 & $6.21 \times 10^{19}$ & 7.61 & 1445.0 & B8 & 1.09 \\
HD 169454 & G017.5385--0.6699 & 0.34 & 0.65 & $2.96 \times 10^{18}$ & 6.20 & 1947.0 & B1 & 0.67 \\
HD 152236 & G343.0276+0.8703 & 0.17 & 0.85 & $3.95 \times 10^{18}$ & 4.49 & 1975.0 & B1.5 & 1.16 \\
HD 152234 & G343.4625+1.2157 & 0.70 & 0.15 & $6.53 \times 10^{17}$ & 5.33 & 1935.0 & B0.5 & 1.18 \\
HD 93129A & G287.4097--0.5738 & 0.50 & 31.71 & $2.23 \times 10^{20}$ & 7.17 & 2424.0 & O3. & 0.18 \\
HD 152408 & G344.0838+1.4914 & 0.71 & 0.53 & $2.06 \times 10^{18}$ & 5.67 & 1796.0 & O8. & 2.15 \\
CD --43 4690 & G264.2096+0.2146 & 0.20 & 0.81 & $4.42 \times 10^{18}$ & 9.16 & 2137.0 & O7.5 & 0.20 \\
CP --45 3218 & G266.1819--0.8482 & 0.37 & 0.29 & $1.21 \times 10^{18}$ & 8.75 & 1878.0 & O9.5 & 2.07 \\
    \bottomrule
    \end{tabular}
    \label{massive-stars}
    
\end{table*}

\begin{table*}
\begin{threeparttable}
    \centering
    \caption{Properties of Wolf Rayet Stars found in SMGPS. The $S_{1.3\text{ GHz}}$ fluxes are from SMGPS whereas $S_{4.8\text{ GHz}}$ and $S_{8.6\text{ GHz}}$ are from either VLA or ATCA.}
    \begin{tabular}{llccrrcllll}
\toprule
\parbox{1.3cm}{Target \\ Name} & 
SMGPS ID & \parbox{0.7cm}{ $\sigma_{\rm SMGPS}$ \\ ($\arcsec$)}  & \parbox{0.8cm}{ $S_{1.3\text{ GHz}}$ \\ (mJy)} & 
\parbox{0.8cm}{ $S_{4.8\text{ GHz}}$ \\ (mJy)} & 
\parbox{0.8cm}{ $S_{8.6\text{ GHz}}$ \\ (mJy)} & 
\parbox{1.4cm}{$L_R$ \\ $(\text{erg}\,\mathrm{s^{-1}}\,\mathrm{Hz^{-1}}
    )$} & 
\parbox{0.8cm}{$G$ \\ (mag)} & 
\parbox{0.8cm}{Distance \\ (pc)} & 
\parbox{0.8cm}{Spectral \\ Type} & 
\parbox{0.8cm}{sep. \\ ($\arcsec$)} \\
\midrule

HD 165763 & G009.2388--0.6132 & 1.18 & 0.26 & 0.33\tnote{(1)} & -- & $5.42 \times 10^{17}$ & 7.50 & 1331 & WC6 & 0.93  \\
HD 165688 & G010.8001+0.3944 & 0.54 & 0.42 & 1.17\tnote{(2)} & 1.77\tnote{(2)} & $1.58 \times 10^{18}$ & 9.21 & 1774 & WN5-6b & 0.10 \\
NaSt1$^{*}$ & G033.9152+0.2638 & 0.26 & 1.83 & -- & -- & $1.43 \times 10^{19}$ & 13.14 & 2557 & WN10 & 0.86 \\
HD 76536 & G267.5519--1.6370 & 0.91 & 0.22 & 0.46\tnote{(3)}  & 0.26\tnote{(3)} & $7.83 \times 10^{17}$ & 8.63 & 1719 & WC & 0.58 \\
HD 79573 & G271.4235--1.0817 & 0.52 & 0.29  & $< 0.33$\tnote{(3)} & 0.69\tnote{(3)} & $2.09 \times 10^{18}$ & 10.16 & 2455 & WC6 & 0.15 \\
HD 151932 & G343.2090+1.4320 & 0.62 & 0.61 & $<2.0$\tnote{(4)} & $\sim 2.0$\tnote{(4)} & $2.14 \times 10^{18}$ & 6.30 & 1706 & WN7h & 1.22 \\
HD 152270 & G343.4873+1.1643 & 0.21 & 0.52 & 0.86\tnote{(5)} & 1.67\tnote{(5)} & $1.25 \times 10^{18}$ & 6.56 & 1426 & WC7+O5-8 & 1.21 \\
HD 152408 & G344.0838+1.4914 & 0.71 & 0.53 & -- & 0.68\tnote{(6)} & $2.06 \times 10^{18}$ & 5.67 & 1796 & O8Iape & 2.15 \\
HD 318016 & G355.2070--0.8719 & 0.35 & 0.28 & 0.94\tnote{(6)} & 1.09\tnote{(6)} & $1.54 \times 10^{18}$ & 10.89 & 2154 & WN8/WC7 & 1.52 \\
\bottomrule
\end{tabular}

\begin{tablenotes}
    \item \textbf{References:} (1) \cite{Bieging:1982}; (2) \cite{Skinner:2002}; (3) \cite{Leitherer:1997}; (4) \cite{SetiaGunawan:2003}; (5) \cite{Montes:2009}; (6) \cite{Cappa:2004}.
\end{tablenotes}
    \label{wolfrayet}

\end{threeparttable}
\end{table*}

\subsubsection{Chromospherically Active Stars}
    As expected, our analysis reveals the detection of a number of chromospherically active stars, which have SMGPS flux density of tenths of mJy, except for FI Cru and V841 Cen, which have fluxes of 1.17 and 4.52 mJy, respectively. See Table \ref{rscn} for the radio and optical properties of these stars. Table \ref{rscn} shows that, within our limits, the RS CVn systems, in particular the ZTF-discovered ones show, on average, a higher radio luminosity, in excess of 10$^{18}$ ergs$^{-1}$ Hz$^{-1}$, than the BY Dra-type systems, with much fainter radio luminosities of only a few times 10$^{16}$ ergs$^{-1}$ Hz$^{-1}$. The luminosity of the ZTF-detected RS CVn's puts them mostly at a luminosity above the main sequence, evolving to the giant branch. We suggest variability may influence these results, as flaring objects are easier to detect in a flux-limited survey such as SMGPS than non-flaring objects when they have the same quiescent luminosity.

\begin{table*}
    \centering
    \caption{Some chromospheric active stars comprising RS CVn and BY Dra  type stars in SMGPS}

\begin{tabular}{llcrlrrll}
\toprule
\parbox{1.3cm}{Target \\ Name} & SMGPS ID & \parbox{0.7cm}{ $\sigma_{\rm SMGPS}$ \\ ($\arcsec$)} & \parbox{0.8cm}{ $S_{1.3\text{ GHz}}$ \\ (mJy)}  & \parbox{1.4cm}{$L_R$ \\ $(\text{erg}\,\mathrm{s^{-1}}\,\mathrm{Hz^{-1}}
    )$} & \parbox{0.8cm}{ $G$ \\ (mag)} & \parbox{0.8cm}{Distance \\ (pc)}  & \parbox{0.8cm}{Object \\ Type} & \parbox{0.8cm}{sep. \\ ($\arcsec$)} \\
\midrule
ZTF J184430.70--050304.7 & G027.6482--0.7654 & 1.08 & 0.21 & $1.90 \times 10^{18}$ & 16.45 & 2759.0 & RS CVn & 1.72 \\
ZTF J191247.50+084858.6 & G043.2043--0.6840 & 0.23 & 0.82 & $1.94 \times 10^{18}$ & 16.51 & 1409.0 & RS CVn & 2.87 \\
ZTF J190554.14+095651.5 & G043.4253+1.3447 & 0.67 & 0.99 & $3.56 \times 10^{17}$ & 17.03 & 547.0 & BY Dra & 2.89 \\
ZTF J191352.11+090822.7 & G043.6149-0.7698 & 0.52 & 0.43 & $9.13 \times 10^{17}$ & 17.84 & 1339.0 & BY Dra & 0.31 \\
ZTF J191426.01+125823.7 & G047.0745+0.8881 & 1.20 & 0.16 & $2.36 \times 10^{17}$ & 15.41 & 1121.0 & RS CVn & 1.56 \\
V353 Sge & G052.7181+0.0199 & 0.38 & 0.42 & $3.53 \times 10^{17}$ & 13.46 & 833.0 & RS CVn & 0.78 \\
ZTF J194301.58+213917.3 & G057.9609--0.9370 & 0.65 & 0.21 & $1.32 \times 10^{18}$ & 15.72 & 2282.0 & RS CVn & 2.85 \\
ZTF J193809.60+233031.7 & G059.0179+0.9545 & 1.23 & 0.18 & $7.92 \times 10^{16}$ & 17.31 & 614.0 & BY Dra & 0.74 \\
ZTF J194739.92+225953.9 & G059.6612--1.1939 & 1.58 & 0.22 & $5.18 \times 10^{17}$ & 18.40 & 1414.0 & BY Dra & 2.39 \\
ZTF J194453.78+243902.4 & G060.7708+0.1844 & 0.51 & 0.28 & $3.97 \times 10^{18}$ & 15.40 & 3448.0 & RS CVn & 2.38 \\
V* IO Vel & G277.0199--1.3793 & 0.13 & 0.48 & $1.71 \times 10^{16}$ & 6.85 & 172.0 & RS CVn & 0.17 \\
V915 Cen & G295.3480--1.9521 & 1.01 & 0.38 & $6.32 \times 10^{16}$ & 7.98 & 374.0 & RS CVn & 0.80 \\
ASAS J115948--6136.2 & G296.8366+0.6524 & 1.14 & 0.07 & $1.10 \times 10^{15}$ & 11.41 & 113.0 & BY Dra & 0.87 \\
FI Cru & G302.1673--0.6680 & 0.06 & 1.17 & $1.74 \times 10^{16}$ & 10.34 & 111.0 & RS CVn & 1.09 \\
V851 Cen & G309.1885+0.8649 & 0.29 & 0.76 & $4.97 \times 10^{15}$ & 7.35 & 74.0 & RS CVn & 1.59 \\
V841 Cen & G315.3003--0.0293 & 0.02 & 4.52 & $4.86 \times 10^{16}$ & 8.38 & 94.0 & RS CVn & 0.37 \\
\bottomrule
\end{tabular}

    \label{rscn}
\end{table*}

\subsubsection{Young Stellar Objects}
The third main group identifiable in Figure\ \ref{hrdiag} are Young Stellar Objects (YSOs). 
Depending on their evolutionary stage,  YSOs can be radio emitters at cm wavelengths as shown by \cite{Anglada:2018}. We found 45 YSOs in our sample that have been previously classified by \cite{Zari:2018}. 
The YSOs had SMGPS fluxes ranging from 0.06 to 1.25 mJy and radio luminosities ranging from $ 10^{15} - 10^{18}$ $\mathrm{erg \, s^{-1} \, Hz^{-1}}$. Some massive YSOs contained in the list are CPD--63\,2367 and 2MASS J12271665--6239142, with spectral types K2 and K3, respectively. The majority of the YSOs are within 450 pc, except for 2MASS J08384049--4044465, CD-39 4570, CD-40 4510, 2MASS J08380902--4020313 and UCAC4 353--125341, which have a distance of 841, 899, 915, 989 and 2617 pc, respectively. As the horizon of our volume is much larger than this, and we are in the Galactic Plane and do not expect a fall-off of YSOs with distance, this indicates that the intrinsic YSO radio luminosity, at 1.3 GHz, is in the range of 10$^{15} - 10^{18}$ ergs$^{-1}$ Hz$^{-1}$, but no brighter, unless our association is limited by a horizon in the optical identification of YSOs. The YSOs in our sample are listed in Table \ref{yso}. The massive young stellar objects obtained from the SMGPS have been reported in \citet{Obonyo:2024}.

\begin{table*}
    \centering
    \caption{Young stellar objects in the SMGPS. The spectral type information is from the Gaia Astrophysical parameter catalogue.}

\begin{tabular}{llcrlrrll}
\toprule
\parbox{1.3cm}{Target \\ Name} & 
SMGPS ID & 
\parbox{0.7cm}{ $\sigma_{\rm SMGPS}$ \\ ($\arcsec$)} &
\parbox{0.8cm}{ $S_{1.3\text{ GHz}}$ \\ (mJy)}  & 
\parbox{1.4cm}{$L_R$ \\ $(\text{erg}\,\mathrm{s^{-1}}\,\mathrm{Hz^{-1}})$} & 
\parbox{0.8cm}{$G$  \\ (mag)} & 
\parbox{0.8cm}{Distance \\ (pc)}  & 
\parbox{1.0cm}{Spectral \\ Type} & 
\parbox{0.8cm}{sep. \\ ($\arcsec$)} \\
\midrule
UCAC4 353--125341 & G011.3979--0.6973 & 0.58 & 0.92 & $7.55 \times 10^{18}$ & 12.46 & 2617 & B & 1.57 \\
UCAC4 378--108000 & G016.1862+0.8533 & 0.72 & 0.37 & $9.93 \times 10^{15}$ & 14.86 & 149 & M & 0.35 \\
Gaia DR3 4156018236362660224 & G021.4124+1.0264 & 0.46 & 0.34 & $1.98 \times 10^{16}$ & 13.53 & 221 & K & 2.47 \\
Gaia DR3 5544982828473420416 & G252.3447--0.7332 & 0.65 & 0.08 & $1.28 \times 10^{16}$ & 17.06 & 356 & M & 2.17 \\
Gaia DR3 5546477816392534912 & G252.4185--0.2970 & 0.87 & 0.06 & $1.46 \times 10^{16}$ & 16.63 & 436 & M & 0.53 \\
Gaia DR3 5546226891521982848 & G253.1317+0.1675 & 1.38 & 0.09 & $1.12 \times 10^{16}$ & 16.88 & 329 & M & 2.08 \\
2MASS J08252089--3636492 & G255.5157+0.7245 & 0.24 & 0.32 & $1.43 \times 10^{18}$ & 19.40 & 1921 & -- & 1.10 \\
Gaia DR3 5541985972093756672 & G255.9115--0.1447 & 0.52 & 0.16 & $3.70 \times 10^{16}$ & 14.90 & 434 & M & 0.95 \\
Gaia DR3 5540111991958358528 & G257.0441--1.2726 & 0.18 & 0.33 & $4.81 \times 10^{16}$ & 13.94 & 346 & K & 1.22 \\
UCAC4 254--026265 & G257.7316--0.9481 & 0.94 & 0.07 & $1.03 \times 10^{16}$ & 14.52 & 343 & M & 0.24 \\
Gaia DR3 5527943421893354880 & G258.3557--1.5406 & 0.25 & 0.27 & $3.74 \times 10^{16}$ & 16.00 & 339 & M & 1.72 \\
Gaia DR3 5527871021626231424 & G258.5264--0.7309 & 0.94 & 0.09 & $1.27 \times 10^{16}$ & 11.86 & 349 & K & 1.83 \\
Gaia DR3 5527887960975776768 & G258.8188--2.0424 & 1.25 & 0.28 & $3.42 \times 10^{16}$ & 16.64 & 320 & M & 1.05 \\
Gaia DR3 5527703109886064768 & G258.9917--1.8043 & 0.23 & 0.44 & $5.90 \times 10^{16}$ & 15.64 & 336 & M & 1.74 \\
CD--39  4570 & G259.6313+0.5741 & 0.52 & 0.11 & $1.06 \times 10^{17}$ & 9.49 & 899 & B & 0.35 \\
2MASS J08380902--4020313 & G260.0035+0.5376 & 0.60 & 0.12 & $1.39 \times 10^{17}$ & 16.77 & 989 & K & 0.66 \\
2MASS J08384049--4044465 & G260.3855+0.3713 & 1.13 & 0.10 & $8.41 \times 10^{16}$ & 16.84 & 841 & M & 0.75 \\
CD--40  4510 & G260.7650+0.6390 & 0.86 & 0.75 & $7.51 \times 10^{17}$ & 10.14 & 915 & B & 0.88 \\
Gaia DR3 5526683209770386048 & G260.8474--0.6347 & 0.46 & 0.17 & $7.39 \times 10^{15}$ & 16.20 & 188 & M & 0.58 \\
Gaia DR3 5523337086650972416 & G262.2781--1.4567 & 0.33 & 0.37 & $4.37 \times 10^{16}$ & 15.31 & 315 & M & 0.45 \\
2MASS J08363137--4327018 & G262.2974--1.5823 & 0.72 & 0.16 & $1.87 \times 10^{16}$ & 16.00 & 315 & M & 0.19 \\
UCAC4 174--040115 & G279.2739--0.8565 & 1.38 & 0.08 & $1.95 \times 10^{15}$ & 13.83 & 139 & M & 1.37 \\
2MASS J10462256--6113379 & G288.4563--1.9209 & 0.86 & 0.30 & $3.39 \times 10^{15}$ & 11.27 & 97 & K & 2.12 \\
2MASS J11094376--6005457 & G290.5718+0.3120 & 0.59 & 1.98 & $2.52 \times 10^{19}$ & 19.06 & 3259 & -- & 1.13 \\
UCAC4 146--088848 & G293.9264+0.6652 & 0.78 & 0.46 & $4.27 \times 10^{15}$ & 13.40 & 87 & M & 0.85 \\
UCAC4 134--060461 & G295.8000--1.2324 & 0.25 & 0.37 & $4.87 \times 10^{15}$ & 13.14 & 104 & M & 0.60 \\
Gaia DR3 5334774553399893376 & G296.0810+0.5246 & 0.91 & 0.08 & $1.42 \times 10^{16}$ & 16.05 & 381 & M & 2.09 \\
UCAC4 141--089534 & G296.7991+0.4022 & 0.67 & 0.10 & $1.41 \times 10^{15}$ & 14.74 & 109 & M & 0.21 \\
HD 104919 & G297.8785--1.7522 & 0.84 & 0.27 & $3.34 \times 10^{15}$ & 9.24 & 100 & G & 0.35 \\
2MASS J12185449--6158348 & G299.1203+0.6543 & 0.34 & 0.22 & $2.88 \times 10^{15}$ & 12.60 & 104 & M & 1.02 \\
FP Cru A & G299.6665--1.3831 & 1.00 & 1.25 & $1.51 \times 10^{16}$ & 10.61 & 100 & K & 1.02 \\
2MASS J12224731--6337572 & G299.7570--0.9375 & 0.22 & 0.41 & $6.59 \times 10^{15}$ & 15.26 & 115 & M & 0.45 \\
UCAC4 134--074521 & G299.7712--0.5818 & 0.13 & 0.55 & $8.43 \times 10^{15}$ & 12.11 & 113 & M & 0.35 \\
UCAC4 135--076980 & G299.8714--0.4793 & 1.04 & 0.10 & $1.24 \times 10^{15}$ & 13.56 & 103 & M & 0.14 \\
UCAC4 135--077407 & G299.9925--0.2776 & 0.95 & 0.11 & $1.49 \times 10^{15}$ & 15.20 & 104 & M & 0.48 \\
2MASS J12271665--6239142 & G300.1608+0.0876 & 0.69 & 0.12 & $1.54 \times 10^{15}$ & 10.49 & 103 & K & 1.35 \\
CD--62   657 & G300.3544--0.5931 & 0.39 & 0.50 & $6.85 \times 10^{15}$ & 9.12 & 107 & K & 0.42 \\
UCAC4 131--071514 & G301.0330--1.0700 & 0.36 & 0.19 & $2.60 \times 10^{15}$ & 14.63 & 107 & M & 0.21 \\
CPD-63  2367 & G301.2970--0.9207 & 0.47 & 0.42 & $5.42 \times 10^{15}$ & 9.73 & 104 & K & 0.18 \\
2MASS J12421136--6403058 & G301.9194--1.1981 & 1.22 & 0.08 & $1.05 \times 10^{15}$ & 15.88 & 103 & M & 2.15 \\
UCAC4 129--071513 & G303.0336--1.4393 & 0.19 & 0.97 & $1.20 \times 10^{16}$ & 14.02 & 101 & M & 1.20 \\
UCAC4 142--109996 & G304.1187+1.1595 & 0.38 & 0.32 & $4.59 \times 10^{15}$ & 14.71 & 109 & M & 1.40 \\
UCAC4 142--113245 & G305.7868+1.0493 & 0.46 & 0.51 & $7.47 \times 10^{15}$ & 12.41 & 110 & M & 2.23 \\
UCAC4 145--145317 & G310.9585+0.8209 & 0.29 & 0.76 & $1.46 \times 10^{16}$ & 12.51 & 126 & M & 1.60 \\
Gaia DR3 5967858367784856448 & G341.9669+1.3341 & 0.51 & 0.31 & $1.92 \times 10^{15}$ & 14.07 & 72 & M & 1.42 \\
2MASS J17275960--3607344 & G351.9424--0.7353 & 0.79 & 0.30 & $5.95 \times 10^{15}$ & 11.57 & 129 & K & 2.61 \\
Gaia DR3 4054971915149637120 & G356.4892--0.4356 & 0.85 & 0.18 & $4.13 \times 10^{16}$ & 16.16 & 440 & K & 2.64 \\
\bottomrule
\end{tabular}

    \label{yso}
\end{table*}

\subsubsection{Other classes}
A few sources in other classes stand out from Figure\ \ref{hrdiag}: \\ \vspace{1mm}
\\
{\bf -- Nearby Stars}\\
Some of the nearby stars in our sample include PM J14111--6155 (Gaia DR3 5866128597052432128), a low-mass dMe star, with spectral type M1Ve, located at a distance of 19.5 pc \citep{Riaz:2006}. The source is a $G = 10$ magnitude star with an SMGPS flux density of 0.27 mJy. It has also been detected using the MeerKAT transient pipeline. For more details of the detection, see Smirnov and Ramailla {\it (in prep.)}. Additionally, UCAC4 196--027069 (Gaia DR3 5313507318402118912) is a nearby star at a distance of 32 pc with a brightness of $G = 13$ mag and an SMGPS flux density of 0.15 mJy. The source has been determined to have an exoplanet in its habitable zone \citep{Kaltenegger:2019}. LP 452--10 (Gaia DR3 4320992440039701120) is a dMe star with an M3.5Ve spectral type and a $V$ magnitude of 13 and radio flux of 0.47 mJy. It is located at a distance of 21 pc \citep{Lepine:2013}. Gaia DR3 5526948535671237888 is an M dwarf star. The star is located at a distance of 48.2 pc. All SMGPS-\textit{Gaia} associations within distances of $\leq 50$ pc are listed in Table \ref{dist50pc}.

Although some of the nearby stars in our sample are classified as high proper motion stars based on their angular motions derived from legacy observations and classifications. To better understand the kinetic properties of these stars, we computed the tangential velocities using \textit{Gaia} DR3 proper motion and parallax information. Most of the stars in our sample have tangential velocities in the range of 21-90 km/s, with only PM J14111-6155 falling below 20 km/s. These velocities indicate our sample is kinematically warm disk stars, possibly associated with a thick disk population. \\ \vspace{1mm}
\\
{\bf -- The novalike Cataclysmic Variable V603 Aql} \\
V603 Aql is an old nova (Nova Aql 1918), now classified as a novalike cataclysmic variable. It has been detected multiple times in the radio with the VLA in the 4–6 and 8–10 GHz frequency bands, with a flux reaching 0.19 mJy \citep{Coppejans:2015, Hewitt:2020}. Also, it was observed with MeerKAT as part of ThunderKAT radio transient survey programme in the $L$ band with flux reaching 0.233 mJy \citep{Hewitt:2020}. In this survey, the source recorded a flux density of 0.36 mJy and a luminosity of $ 4.36 \times 10^{16}\mathrm{erg \, s^{-1} \, Hz^{-1}}$. V603 Aql has significant H$\alpha$ emission from \textit{Gaia} spectroscopy, expected for an accretion-driven CV. The radio emission from the novalike CV is believed to be produced by free-free thermal radiation powered by the white dwarf or non-thermal radiation from additional emission sources such as shocks \citep{Coppejans:2015, Gulati:2023}. \\ \vspace{1mm}
\\
{\bf -- H$\alpha$ emission stars}\\
While many of the previous classes of stellar radio emitters exhibit $\text{H}\alpha$ emission lines, in this section, we present additional examples.
Balmer emission lines are indicators of diffuse gas of intermediate temperatures and are a tracer of stellar magnetic activity and hence, coronal and chromospheric activity in the stellar atmospheres \citep{Newton:2017, Traven:2015}.

\begin{figure}
	\includegraphics[width=\columnwidth]{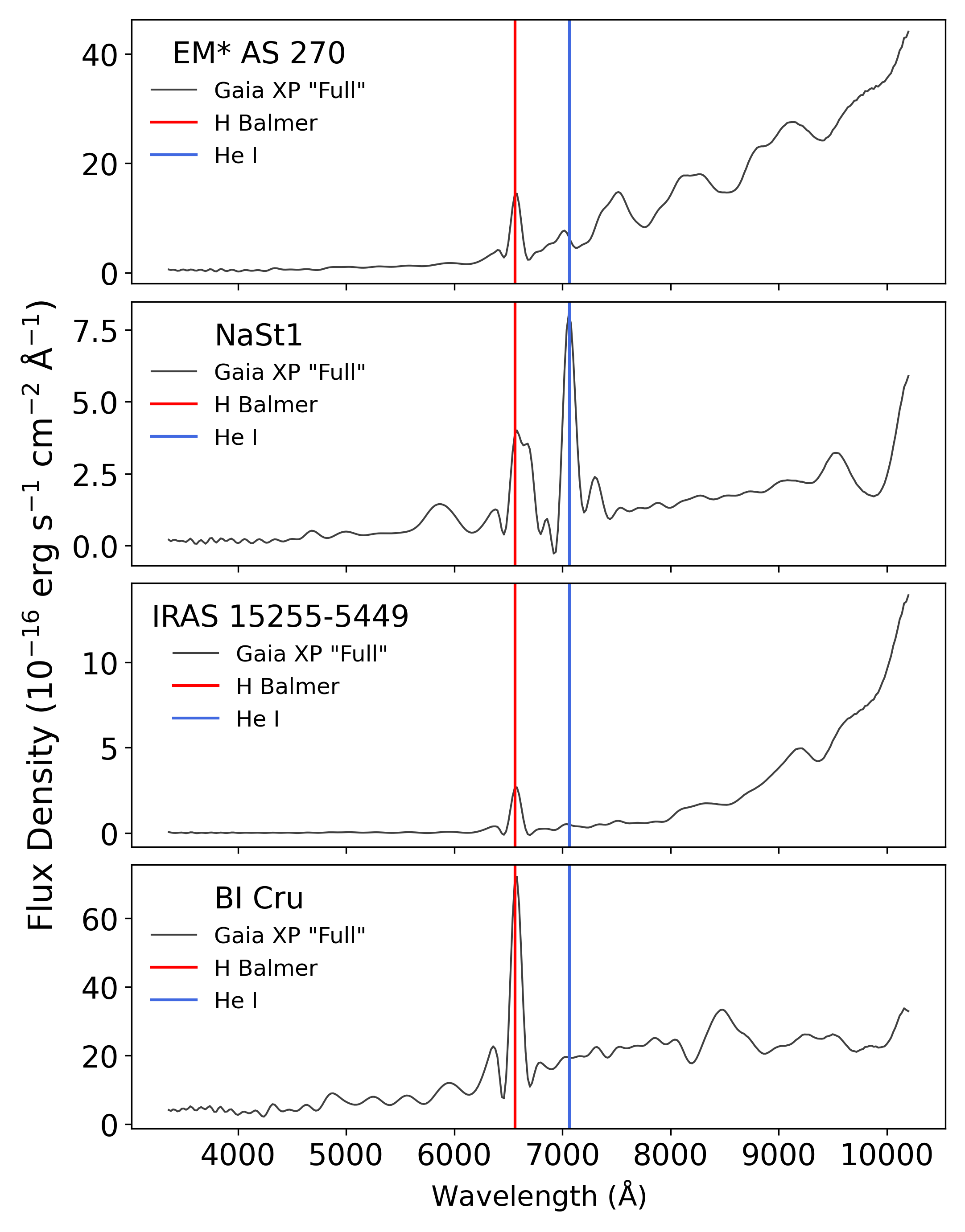}
    \caption{Gaia low-resolution spectra of SMGPS - Gaia candidates  (EM* AS 270, NaSt1, IRAS 15255-5449 and BI Cru) with H$\alpha$ EW < $-3$ nm. NaSt1 spectrum shows strong He I emission, which is a common line for Wolf-Rayet stars.}
    \label{gaisapec}
\end{figure}

The \textit{Gaia} mission performed a spectroscopic survey for sources with $G$ < 17.6 mag. We searched the Astrophysical catalogue of \textit{Gaia} DR3 and found sources in our sample with H$\alpha$ emission, which include BI Cru, EM* AS 270, LS IV +00 5 and IRAS 15255--5449, with H$\alpha$ equivalent width (EW) < $-$3 nm. All stars with significant H$\alpha$ lines (EW < $-$1 nm) are listed in Table \ref{halpha}. These comprise 5 Wolf-Rayet stars, 2 symbiotic stars, 1 long-period variable, 1 AGB candidate, and 1 YSO. The low-resolution spectra of EM* AS 270, NaSt1 (LS IV +00 5), IRAS 15255--5449 and BI Cru containing $\text{H} \alpha$ emission lines are shown in Figure \ref{gaisapec}. \\ \vspace{1mm}
\\
{\bf -- White Dwarfs}\\
We have found some candidate white dwarfs in our cross-matching exercise. This presents an opportunity to explore the less well-understood radio characteristics of white dwarfs. Although white dwarfs are typically not strong radio emitters, certain conditions can lead to detectable radio emissions. A recent follow-up of white dwarfs from a VLASS and \textit{Gaia} crossmatch at 3 GHz, single white dwarfs seen as radio emitters \citep{Pelisoli:2024}. In some close binary systems containing a white dwarf, radio emission has been detected, likely due to the interaction of the white dwarf's magnetic field with its companion \citep{Stanway:2016, Barrett:2020, Pelisoli:2024}.

\cite{Pritchard:2024} identified two cataclysmic variables comprising white dwarfs through a circular polarisation search with ASKAP at 887.5 MHz within 200 pc. Hence, we argue that the white dwarfs in our case may likely fall into specific categories that explain their radio emission. Some may be magnetic white dwarfs, which can emit radio waves through cyclotron maser emission, particularly if they are part of magnetic cataclysmic variables such as polars or intermediate polars where a mass transfer from a companion star is interacting with the white dwarf's magnetic field and leading to radio flaring or persistent emission \citep{Bastian:1988, Ferrario:2015, Marsh:2016, Buckley:2017, Barrett:2020}. On the other hand, a few may belong to binary systems where the white dwarf is accreting material from a companion, a scenario often seen in symbiotic binaries or novae \citep{Hewitt:2020, Gulati:2023}. In these cases, the accretion process and the interaction with the white dwarf’s magnetic field could produce radio emissions. \\ \vspace{1mm}
\\
{\bf -- Red Dwarfs} \\
We found a significant number of main-sequence red dwarfs in our sample. This is evident in Figure \ref{hrdiag} where they dominate the lower end of the main sequence population, with a median radio luminosity of $ 8.9 \times 10^{15}\mathrm{erg \, s^{-1} \, Hz^{-1}}$ and all within a distance of less than 300 pc. While they are not typically strong radio emitters, certain conditions can result in significant radio activity. Radio emission from red dwarfs is generally associated with their magnetic activity. These stars often exhibit strong magnetic fields and can generate radio waves through gyrosynchrotron radiation or coherent processes, such as electron cyclotron maser emission \citep{Vedantham:2020}. This is especially pronounced in flare stars, a subset of red dwarfs known for their intense and sudden increases in brightness due to magnetic reconnection events \citep{White:1986}. Hence, the red dwarfs identified in our study likely represent stars with heightened magnetic activity, either as flare stars or through persistent magnetic phenomena.\\ \vspace{1mm}
\\ 
{\bf -- Red Dwarf - White Dwarf Gap}\\
We refer to the sparsely populated sources in the valley between the main and the
white dwarf sequences of the Hertzsprung-Russell Diagram shown in Figure \ref{hrdiag}, comprising the cool, low luminosity red dwarfs and the hot, compact white dwarfs. To investigate the nature of the sources in the region, we examined the astrometric consistencies of the sources by comparing the \textit{Gaia} DR2 and DR3 proper motion and calculated the significance of the difference using the combined uncertainties from both releases. Fourteen sources exhibit statistically significant changes in proper motion at more than $3\sigma$ level. These are likely red dwarf-quasar blends or astrometrically problematic sources. In addition, 17 sources did not have a \textit{Gaia} DR2 proper motion determination, and are only present in DR3. Therefore, a comparison cannot be made. This analysis suggests that a fraction of our sample in the red dwarf-white dwarf gap may be spurious blends, but a number of genuine gap objects are present.

\begin{table*}
\centering
\caption{Stars within 50 pc.}
\begin{tabular}{llclrlrrrrrc}

\toprule
\parbox{1.3cm}{Target \\ Name} & 
SMGPS ID & \parbox{0.7cm}{ $\sigma_{\rm SMGPS}$ \\ ($\arcsec$)} & \parbox{0.7cm}{ $S_{\mathrm{1.3 GHz}}$ \\ (mJy)} & 
\parbox{1.4cm}{$L_R$ \\ $(\text{erg}\,\mathrm{s^{-1}}\,\mathrm{Hz^{-1}}
    )$} & 
\parbox{0.8cm}{$G$ \\ (mag)} & 
\parbox{1.0 cm}{PM \\ $(\text{mas}\ \rm yr^{-1})$} & 
\parbox{0.7 cm}{Distance \\ (pc)} &
\parbox{1.0 cm}{Spectral \\ Type} & 
\parbox{0.7 cm}{$\text{V}_t$\\ (km/s)} & 
\parbox{0.8cm}{sep. \\ ($\arcsec$)} \\
\midrule
LP  452--10 & G051.7852+0.0045 & 0.51 & 0.50 & $2.62 \times 10^{14}$ & 11.82 & 210.36 & 20 & M & 20.9 & 1.94 \\
L  459-76 & G258.7038+0.6793 & 0.15 & 0.44 & $1.00 \times 10^{15}$ & 11.13 & 238.83 & 43 & K & 49.3 & 0.80 \\
Gaia DR3 5526948535671237632 & G260.1206--0.4461 & 0.41 & 0.16 & $4.36 \times 10^{14}$ & 14.53 & 133.70 & 48 & M & 30.6 & 0.31 \\
UCAC4 196--027069 & G272.3193--1.4081 & 0.48 & 0.15 & $1.84 \times 10^{14}$ & 13.03 & 144.62 & 31 & M & 21.9 & 0.33 \\
Gaia DR3 6054608701362026496 & G299.7899--0.0183 & 1.62 & 0.32 & $9.70 \times 10^{14}$ & 12.59 & 206.22 & 49 & M & 48.8 & 2.01 \\
PM J14111--6155 & G312.1615--0.4949 & 0.66 & 0.28 & $1.28 \times 10^{14}$ & 10.06 & 161.26 & 19 & M & 14.9 & 0.90 \\
2MASS J14500938--6036019 & G317.0007--1.0135 & 1.28 & 0.25 & $4.17 \times 10^{14}$ & 14.31 & 197.76 & 37 & M & 35.2 & 1.13 \\
UCAC4 261-113470 & G348.5139+1.2131 & 0.52 & 0.26 & $6.30 \times 10^{14}$ & 14.03 & 410.72 & 45 & M & 88.4 & 2.37 \\
\bottomrule
\end{tabular}
\label{dist50pc}
\end{table*}

\begin{table*}
    \centering
    \caption{SMGPS - Gaia associations with strong H$\alpha$ Emission Lines.}
    \begin{tabular}{llcrlrllll}
\toprule
\parbox{1.3cm}{ Target \\ Name} & 
SMGPS ID & \parbox{0.7cm}{ $\sigma_{\rm SMGPS}$ \\ ($\arcsec$)} &
\parbox{1 cm}{$S_{\mathrm{1.3 GHz}}$ \\ (mJy) } & 
\parbox{1.4cm}{$L_R$ \\ $(\text{erg}\,\mathrm{s^{-1}}\,\mathrm{Hz^{-1}})$} &
\parbox{0.8cm}{$G$ \\ (mag)} &  
\parbox{0.9cm}{H$\alpha$ EW \\ (nm)} & 
\parbox{1 cm}{ Distance \\ (pc)} & 
\parbox{1.0cm}{Object \\ Type} & 
\parbox{0.8cm}{sep. \\ ($\arcsec$)} \\
\midrule
EM* AS  270 & G009.7016+0.4240 & 1.13 & 0.15 & $1.07 \times 10^{18}$ & 11.27 & --5.96 & 2471 & Symbiotic & 0.57 \\
HD 165688 & G010.8001+0.3944 & 0.54 & 0.42 & $1.58 \times 10^{18}$ & 9.21 & --2.84 & 1774 & Wolf-Rayet & 0.10 \\
NaSt1 & G033.9152+0.2638 & 0.26 & 1.83 & $1.43 \times 10^{19}$ & 13.14 & --3.67 & 2557 & Wolf-Rayet & 0.86 \\
OH 043.6--00.5 & G043.6388--0.4548 & 0.50 & 0.51 & $5.75 \times 10^{18}$ & 17.21 & --2.07 & 3062 & AGB Cand. & 0.35 \\
HD  76536 & G267.5519--1.6370 & 0.91 & 0.22 & $7.83 \times 10^{17}$ & 8.63 & --1.69 & 1719 & Wolf-Rayet & 0.58 \\
HD  79573 & G271.4235--1.0817 & 0.52 & 0.29 & $2.09 \times 10^{18}$ & 10.16 & --1.53 & 2455 & Wolf-Rayet & 0.15 \\
BI Cru & G299.7201+0.0592 & 0.05 & 4.58 & $4.85 \times 10^{19}$ & 10.46 & --7.64 & 2974 & Symbiotic & 1.05 \\
2MASS J12421136--6403058 & G301.9194--1.1981 & 1.22 & 0.08 & $1.05 \times 10^{15}$ & 15.88 & --1.22 & 103 & YSO & 2.15 \\
PSR B1259--63 & G304.1832--0.9916 & 0.04 & 4.10 & $2.30 \times 10^{19}$ & 9.63 & --1.96 & 2167 & GR binary & 1.42 \\
IRAS 15255-5449 & G324.3375+1.1779 & 0.52 & 1.91 & $1.51 \times 10^{19}$ & 13.73 & --14.48 & 2570 & LPV Cand. & 0.94 \\
HD 152270 & G343.4873+1.1643 & 0.21 & 0.52 & $1.25 \times 10^{18}$ & 6.56 & --1.36 & 1426 & Wolf-Rayet & 1.21 \\
\bottomrule
\end{tabular}
    \label{halpha}
\end{table*}

\subsection{Radio - Optical Flux Relation}
The radio-optical flux relation has been extensively studied in the context of transients and other variable sources and is commonly used as a diagnostic tool to understand the nature of objects that emit radiation in radio and optical wavelengths. One example demonstrating the power of this diagnostic is presented in \cite{Stewart:2018}, which used $1-10$ GHz VLA radio flux density, and  optical SDSS data as well as previously reported radio stellar data from \cite{Wendker:1995} to investigate the relationship between the radio $(F_{r})$ and optical $(F_{o})$ flux densities of different classes of compact radio sources. The stellar sources comprise RS CVn, Algol-type, double, magnetic, BY Dra-type, Symbiotic, Herbig Wolf-Rayet, T Tauri and high proper-motion stars. The $F_{r} - F_{o}$ correlation can also be used as a diagnostic tool to infer the classes of radio transients, whether they are stellar, non-stellar, or whether they are Galactic or extra-galactic. 
 
\begin{figure*}
	\includegraphics[width=\textwidth]{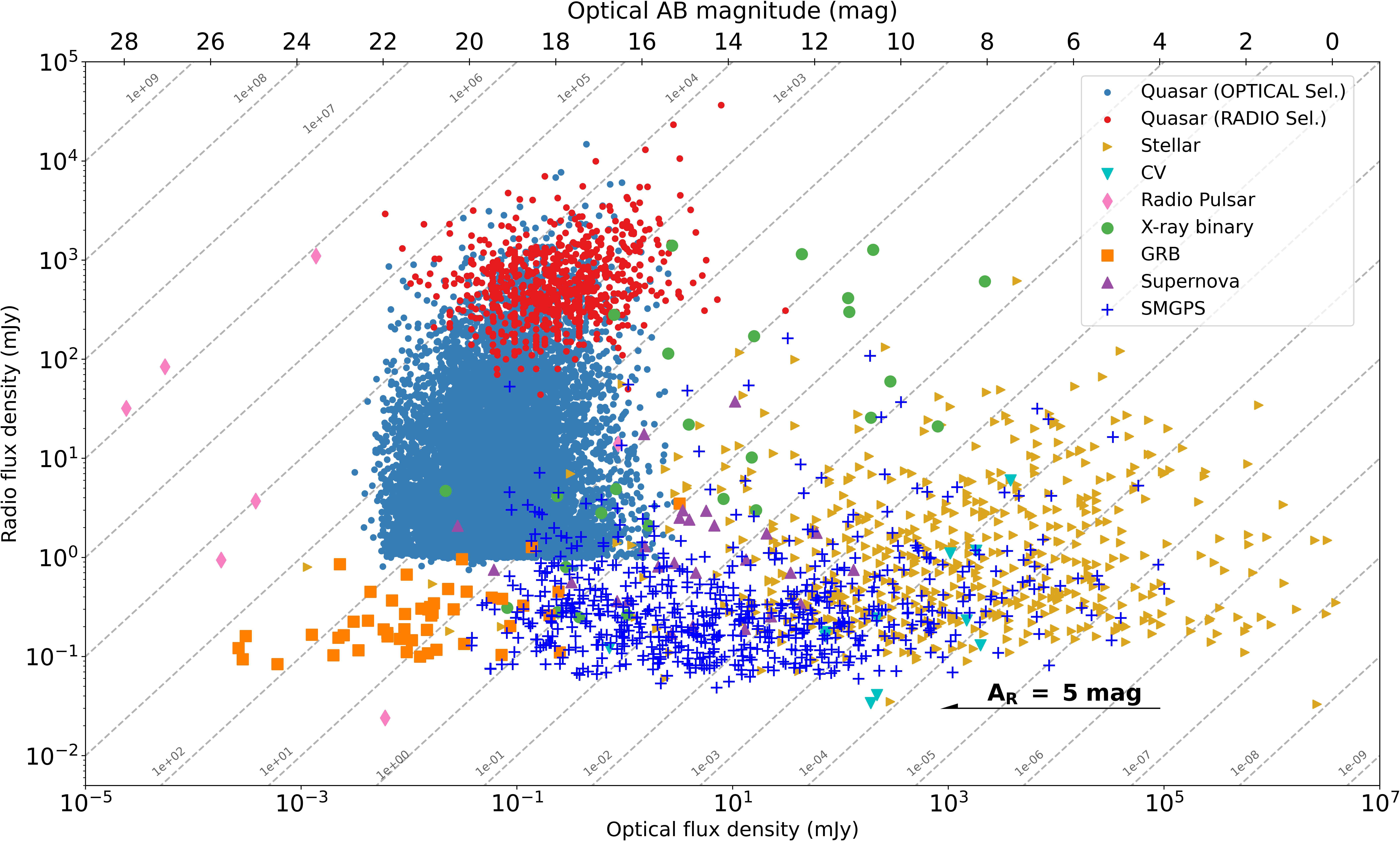}
    \caption{Radio flux $F_{r}$ vs. Optical flux $F_{o}$ parameter space from \citet{Stewart:2018} showing different radio transients. The plot shows different data points from multiple surveys. The markers in the top right legend of the plot represent the data points from VLA, SDSS and other stellar sources with objects ranging from quasars, stellar, cataclysmic variable (CV), radio pulsar, X-ray binary, Gamma-ray burst (GRB) and Supernova. The SMGPS-\textit{Gaia} stellar sources are represented with blue cross markers. The diagonal lines represent constant $F_{r}:F_{o}$ ratios, whose values are labelled. The reddening vector for A$_R = 5$ mag is also shown.
 }
    \label{stewart}
\end{figure*}
In comparing our sample with \cite{Stewart:2018}, we overplotted the optical \textit{Gaia} (extinction corrected) and SMGPS fluxes in the $F_{r} - F_{o}$ parameter space of Figure 1 of \cite{Stewart:2018}. This is shown in Figure \ref{stewart}. Our sources (blue crosses) extend $\sim$1 dex below the VLA FIRST radio limit, to $\sim$0.1 mJy, while covering a range of $\sim$5 dex in optical flux density, from $0.1-10^5$ mJy. This spans stellar regions of the Stewart diagram that include stars, X-ray binaries and cataclysmic variables, and overlaps with extra-galactic sources like GRBs, supernovae and optically selected quasars (the latter expected from extrapolation of the Stewart diagram to fainter radio fluxes). Based on our results, we argue that our SMGPS sample are of typically fainter radio sources than seen in previous surveys and are consistent with stellar radio emitters within the Milky Way Galaxy, given their measured Gaia parallaxes and proper motions. The radio flux densities of the SMGPS sources lie within the expected $5\sigma$ MeerKAT limit, as predicted in Figure 2 of \cite{Stewart:2018}. The faint sources in our SMGPS sample highlight one of the key advantages of MeerKAT: its ability to probe deeper into the faint radio sky. This capability enables us to detect faint radio stellar sources that would have been impossible to identify with earlier instruments.

\section{Conclusions}
The SARAO MeerKAT Galactic Plane Survey has provided a high-sensitivity and deep continuum survey of the Galactic Plane. This work has used the continuum compact sources catalogue derived from SMGPS in combination with other surveys, such as \textit{Gaia} and AllWISE to find radio-emitting stellar candidates. In total, 629 stellar candidates were obtained from this work as reliable radio-emitting stars. This is by far the largest sample of radio stars detected in the Galactic Plane to date. 

The optical colour-magnitude diagram of the radio stellar candidates shows that the stellar sources lie predominantly within, or close to, the main-sequence region of the HR diagram (Fig. \ref{hrdiag}). A literature search of the sources shows that they belong to many known radio star populations, such as binaries, late-type single stars, white dwarfs and red dwarfs.

The optical-radio flux relation can be strongly influenced by variability. In the radio, M dwarfs can show variability by factors of up to 10 and even higher during stellar flares, see e.g. \citet{Quiroga-Nunez:2020, Plant:2024}. In \citet{Collier:1982, Dulk:1985, Garcia-Sanchez:2003}, chromospherically active binaries, like RS CVns, also exhibit changes by several factors in flux. The fraction of each population seen at a given epoch in the radio will depend on the amplitude of variation as well as the duty cycle of the variability. A multi-epoch radio study will be required to further determine these factors.

The more luminous objects identified as radio stars include OB-type stars, Wolf-Rayet stars and blue super-giants. In addition, a sizable number of previously classified coronally or chromospherically active systems (e.g. RS CVn, BY Dra) were also found in the cross-matching at low luminosities. A significant number of young stellar objects (YSOs) and high proper motion sources were also found in the sample.

Our detections demonstrate that MeerKAT is a very sensitive instrument for identifying faint radio stellar sources. Crowding and high stellar density in the Galactic Plane present challenges when cross-matching SMGPS with \textit{Gaia} and other surveys. However, by combining SMGPS data with observations across optical and other wavelengths, we are confident that more radio-emitting stellar candidates in our Galaxy will be discovered. This multi-wavelength approach not only enhances our understanding of these sources but also helps to distinguish genuine radio stars from other radio emitting celestial objects. To further validate these radio star identifications, follow-up radio observations are required, focusing on both polarization studies and additional continuum spectral index measurements. Likewise, optical and X-ray follow-up observations can also provide evidence of a radio star classification through the detection of chromospheric and coronal activity. Results from such studies of sources discovered in this paper will be discussed in two forthcoming publications. These follow-ups provide crucial evidence to confirm that the identified sources are indeed radio stars and will offer an increase in the population of radio-emitting stars in our Galaxy.

\section*{Acknowledgements}

We thank the reviewer for their constructive comments, which helped to improve the clarity and sharpness of the paper.

The authors would like to thank Laura Driessen for her suggestions on the crossmatch process and analysis used in this paper.

ODE would like to thank Cesca Figueras, Merce Romero and Michael Weiler
for helpful discussions about the Gaia extinction correction and the Gaia low-resolution spectra.

This work is based on research fully supported by the National Research Foundation's (NRF) South African Astronomical Observatory (SAAO) PhD Prize Scholarship. PJG is supported by NRF SARChI grant 111692. DAHB acknowledges research support by the National Research Foundation.

The MeerKAT telescope is operated by the South African Radio Astronomy Observatory, which is a facility of the National Research Foundation, an agency of the Department of Science and Innovation.

This work has made use of data from the European Space Agency (ESA) mission
\href{https://www.cosmos.esa.int/Gaia}{\it Gaia}, processed by the {\it Gaia} Data Processing and Analysis Consortium \href{https://www.cosmos.esa.int/web/Gaia/dpac/consortium}{(DPAC)}. Funding for the DPAC has been provided by national institutions, in particular the institutions participating in the {\it Gaia} Multilateral Agreement.

This work is based on data from eROSITA, the soft X-ray instrument aboard SRG, a joint Russian-German science mission supported by the Russian Space Agency (Roskosmos), in the interests of the Russian Academy of Sciences represented by its Space Research Institute (IKI), and the Deutsches Zentrum für Luft- und Raumfahrt (DLR). The SRG spacecraft was built by Lavochkin Association (NPOL) and its subcontractors, and is operated by NPOL with support from the Max Planck Institute for Extraterrestrial Physics (MPE). The development and construction of the eROSITA X-ray instrument was led by MPE, with contributions from the Dr. Karl Remeis Observatory Bamberg \& ECAP (FAU Erlangen-Nuernberg), the University of Hamburg Observatory, the Leibniz Institute for Astrophysics Potsdam (AIP), and the Institute for Astronomy and Astrophysics of the University of Tübingen, with the support of DLR and the Max Planck Society. The Argelander Institute for Astronomy of the University of Bonn and the Ludwig Maximilians Universität Munich also participated in the science preparation for eROSITA.

This publication makes use of data products based on observations obtained with XMM-Newton, an ESA science mission with instruments and contributions directly funded by
ESA Member States and NASA.

This research made use of Astropy,\footnote{http://www.astropy.org} a community-developed core Python package for Astronomy \citep{astropy:2013, astropy:2018}.

This publication makes use of data products from the Two Micron All Sky Survey, which is a joint project of the University of Massachusetts and the Infrared Processing and Analysis Center/California Institute of Technology, funded by the National Aeronautics and Space Administration and the National Science Foundation.

This publication makes use of data products from the Wide-field Infrared Survey Explorer, which is a joint project of the University of California, Los Angeles, and the Jet Propulsion Laboratory/California Institute of Technology, and NEOWISE, which is a project of the Jet Propulsion Laboratory/California Institute of Technology. WISE and NEOWISE are funded by the National Aeronautics and Space Administration.

This work has made use of the Cube Analysis and Rendering Tool for Astronomy (CARTA; \citealt{Comrie:2021}). This research made use of APLpy, an open-source plotting package for Python \citep{aplpy2012, aplpy2019}.

\section*{Data Availability}
Information about the SARAO Galactic Plane Survey can be found in \cite{Goedhart:2024}. The Gaia DR3 data used in this work are available in the CDS Vizier database; the details can be found in \cite{ GaiaCollaboration:2023}. The radio star catalogue from this work will be made available at the CDS (Centre de Données astronomiques de Strasbourg) via anonymous ftp to \href{http://cdsarc.u-strasbg.fr}{cdsarc.u-strasbg.fr (130.79.128.5)}  or
via \href{https://cdsarc.cds.unistra.fr/viz-bin/cat/J/MNRAS}{https://cdsarc.cds.unistra.fr/viz-bin/cat/J/MNRAS} upon publication, and the column description is listed in \ref{appendix1}.




\bibliographystyle{mnras}
\bibliography{example} 



\appendix

\section{SMGPS Radio Star Catalog Column Description}

\begin{table*}
\caption{SMGPS Radio Star Candidates Column Description}
\label{tab:column_desc}
\begin{tabular}{|l|l|}
\hline
Column Name & Column description \\
\hline
mkid & SMGPS identifier \\
fileName & SMGPS image file name \\
lon & Galactic Longitude in deg \\
err\_lon & source-finding fitting error on longitude in deg\\
lat & Galactic latitude in deg \\
err\_lat & source-finding fitting error on latitude in deg \\
peak\_flux & Peak flux density in mJy/beam \\
err\_peak\_flux & source-finding fitting error on peak flux density in mJy/beam \\
int\_flux & Integrated flux density in mJy \\
err\_int\_flux & source-finding fitting error on integrated flux density in mJy \\
a & fitted semi-major axis in arcsec \\
err\_a & error on fitted semi-major axis in arcsec \\
b & fitted semi-minor axis in arcsec \\
err\_b & error on fitted semi-minor axis in arcsec \\
ra & SMGPS J2019.4 right ascension in deg \\
dec & SMGPS J2019.4 declination in deg \\
ra\_err & Error in right ascension in arcsec \\
dec\_err & Error in declination in arcsec \\
radec\_err & Error in both right ascension and declination in arcsec\\
Source & Unique Gaia DR3 Source identifier \\
RA\_ICRS & Right ascension (ICRS) at Epoch=2016.0 in deg \\
DE\_ICRS & Declination (ICRS) at Epoch=2016.0 in deg \\
e\_RA\_ICRS & Standard error of right ascension in mas \\
e\_DE\_ICRS & Standard error of declination in mas \\
SolID & Gaia DR3 Solution identifier \\
Plx & Parallax in mas \\
PM & Total Proper motion in mas/yr \\
Gmag & G-band mean magnitude (mag) \\
BPmag & Integrated BP mean magnitude (mag) \\
RPmag & Integrated RP mean magnitude (mag) \\
RAJ2000 & right ascension (J2000) in deg \\
DEJ2000 & declination (J2000) in deg \\
e\_RAJ2000 & Error in right ascension (J2000) in mas \\
e\_DEJ2000 & Error in declination (J2000) in mas \\
RA2019 & Gaia right Ascension in J2019.4 in deg \\
DEC2019 & Gaia declination in J2019 in deg \\
dRA2019 & Error in right ascension (J2019) in mas \\
dDec2019 & Error in declination (J2019) in mas \\
TIC & TESS Input Catalog identifier \\
AllWISE & AllWISE identifier \\
\_2MASS & 2MASS identifier \\
rpgeo & Median of the photo-geometric distance posterior (pc) \\
fo1sig & normalised offsets between SMGPS and Gaia in arcsec \\ 
so1sig & Confidence level between SMGPS and Gaia match (unitless) \\
reddening & Extinction value in mag \\
Gdmag & Dereddened Gaia magnitude in mag \\
mgg & Absolute magnitude derived from Gdmag in mag \\
opt\_flx\_mj & Gaia Flux in mJy derived from Gdmag \\

o\_lum & Optical luminosity in erg/s \\
r\_lum & Radio luminosity in erg/s/Hz \\
sep\_arcsec & SMGPS - Gaia separation in arcsec \\
\hline
\end{tabular}
\label{appendix1}
\end{table*}


\bsp	
\label{lastpage}
\end{document}